\documentclass[12pt,preprint]{aastex}

\newcommand{\beq}{\begin{equation}}
\newcommand{\eeq}{\end{equation}}

\newcommand{\beqa}{\begin{eqnarray}}
\newcommand{\eeqa}{\end{eqnarray}}

\def\fun#1#2{\lower3.6pt\vbox{\baselineskip0pt\lineskip.9pt
  \ialign{$\mathsurround=0pt#1\hfil##\hfil$\crcr#2\crcr\sim\crcr}}}

\begin{document}

\title{Measuring Baryon Acoustic Oscillations along the line of sight with 
photometric redshifts: the PAU survey}

\shorttitle{The PAU Survey}
\shortauthors{Ben\'{\i}tez et al.}

\author{
N.~Ben\'{\i}tez,\altaffilmark{1} 
E.~Gazta\~naga,\altaffilmark{2} 
R.~Miquel,\altaffilmark{3,4}
F.~Castander,\altaffilmark{2}
M.~Moles,\altaffilmark{5} 
M.~Crocce,\altaffilmark{2}
A.~Fern\'andez-Soto,\altaffilmark{6,7} 
P.~Fosalba,\altaffilmark{2} 
F.~Ballesteros,\altaffilmark{8}
J.~Campa,\altaffilmark{9}
L.~Cardiel-Sas,\altaffilmark{4}
J.~Castilla,\altaffilmark{9} 
D.~Crist\'obal-Hornillos,\altaffilmark{5} 
M.~Delfino,\altaffilmark{11}
E.~Fern\'andez,\altaffilmark{4}\footnote{PAU Coordinator. 
E-mail: {\tt Enrique.Fernandez@ifae.es}} \,
C.~Fern\'andez-Sopuerta,\altaffilmark{2}
J.~Garc\'{\i}a-Bellido,\altaffilmark{10}
J.A.~Lobo,\altaffilmark{2}
V.J.~Mart\'{\i}nez,\altaffilmark{8} 
A.~Ortiz,\altaffilmark{8}
A.~ Pacheco,\altaffilmark{4,11}
S.~Paredes,\altaffilmark{8}\footnote{Permanent address: 
Universidad Polit\'ecnica de Cartagena} \,
M.J.~Pons-Border\'{\i}a,\altaffilmark{8}\footnote{Permanent address: 
Universidad Complutense de Madrid} \,
E.~S\'anchez,\altaffilmark{9}
S.F.~S\'anchez,\altaffilmark{12}
J.~Varela,\altaffilmark{5}
J.F.~de Vicente\,\altaffilmark{9}
}

\altaffiltext{1}{Instituto de Matem\'aticas y F\'{\i}sica Fundamental (CSIC), Madrid}
\altaffiltext{2}{Institut de Ci\`encies de l'Espai (IEEC-CSIC), Barcelona}
\altaffiltext{3}{Instituci\'o Catalana de Recerca i Estudis Avan\c{c}ats, Barcelona}
\altaffiltext{4}{Institut de F\'{\i}sica d'Altes Energies, Barcelona}
\altaffiltext{5}{Instituto de Astrof\'{\i}sica de Andaluc\'{\i}a (CSIC), Granada}
\altaffiltext{6}{Departament d'Astronomia i Astrof\'{\i}sica, Universitat de Val\`encia}
\altaffiltext{7}{Instituto de F\'{\i}sica de Cantabria (CSIC), Santander}
\altaffiltext{8}{Observatori Astron\`omic de la Universitat de Val\`encia}
\altaffiltext{9}{Centro de Investigaciones En\'ergeticas, Medioambientales y Tecnol\'ogicas, Madrid}
\altaffiltext{10}{Instituto de F\'{\i}sica Te\'orica (UAM-CSIC), Madrid}
\altaffiltext{11}{Port d'Informaci\'o Cient\'{\i}fica, Barcelona}
\altaffiltext{12}{Centro Astron\'omico Hispano Alem\'an (CSIC/MPG), Calar Alto}
             


\begin{abstract}
Baryon Acoustic Oscillations (BAO) 
provide a ``standard ruler'' of known physical length, 
making it one of the most promising probes of the nature of dark energy.
The detection of BAO as an excess of power in the galaxy distribution at a certain scale
requires measuring galaxy positions and redshifts. ``Transversal'' 
(or ``angular'') BAO measure the angular
size of this scale projected in the sky and provide information about the angular 
distance. ``Line-of-sight'' (or ``radial'') BAO require very precise redshifts, but provide a direct measurement of 
the Hubble parameter at different redshifts, a more sensitive probe of dark energy.
The main goal of this paper is to show that
it is possible to obtain photometric redshifts with
enough precision ($\sigma_z$) to measure BAO along the line of sight.
There is a fundamental limitation as to how much one can improve the BAO
measurement by reducing  $\sigma_z$. We show that $\sigma_z\sim 0.003 (1+z)$
is sufficient:
a much better precision will produce an oversampling of the BAO peak without
a significant improvement on its detection, while a much worse precision
will result in the effective loss of the radial information.
This precision in redshift can be achieved for
bright, red galaxies,
featuring a prominent $4000$~\AA\  break,
by using a filter system comprising about 40 filters, each with a width
close to $100$~\AA, covering
the wavelength range from $\sim4000$~\AA\  to $\sim8000$~\AA, supplemented
by two broad-band filters similar to the SDSS $u$ and $z$ bands.
We describe the practical implementation of this idea, a new galaxy
survey project, PAU\footnote{Physics of the
Accelerating Universe (PAU): {\tt http://www.ice.cat/pau}},
to be carried out with a telescope/camera combination with an {\it etendue} about 20~m$^2$~deg$^2$,
equivalent to 
a 2~m telescope equipped
with a 6~deg$^{2}$-FoV camera, 
and covering 8000~sq.~deg.\ in the sky in four years. We expect to measure
positions and redshifts for over 14 million red, early-type galaxies
with $L>L_\star$ and $i_{AB}\lesssim 22.5$ in the redshift interval
$0.1<z<0.9$, with a precision $\sigma_z < 0.003(1+z)$. This population
has a number density $n\gtrsim 10^{-3}$ $\textrm{Mpc}^{-3}\textrm{h}^{3}$
galaxies within the 9~$\textrm{(Gpc/h)}^{3}$ volume 
to be sampled by our survey,
ensuring that the error in the determination of the BAO scale is
not limited by shot-noise.
By itself, such a survey will deliver precisions of order $5\%$ in
the dark-energy equation of state parameter $w$, if assumed constant,
and can determine its time derivative when combined with future CMB
measurements. 
In addition, PAU will yield high-quality redshift and low-resolution spectroscopy
for hundreds of millions of other galaxies, including
a very significant high-redshift population. 
The data set produced by this survey will have a unique legacy value, allowing
a wide range of astrophysical studies.
\end{abstract}


\section{Introduction}\label{sec:intro}

Physical Cosmology has recently entered the precision era. This transition has been propelled by the gathering, over the past decade, of unprecedented high-precision data sets for several cosmological observables. The combined analysis of the Cosmic Microwave Background (CMB) anisotropies (e.g.,
\citet{wmap3ext, wmap5ext, boomerang, cbi, acbar, vsa}) with distance-scale measurements at increasingly higher redshifts (e.g.,
\citet{SNe1, astier06, SNe2}) and probes of large-scale structure
\citep{2df, sdss1, sdss2, Tegmark06, hutsi-2} 
      yields a remarkably consistent picture: a spatially flat universe that has started a phase of acceleration of the expansion at the present epoch. From the observations gathered so far this acceleration is consistent with the effect of a cosmological constant, but it may also be caused by the presence of a dynamical energy component with negative pressure, now termed Dark Energy (DE), or might also point to a fundamental modification of our description of gravity. The answer to what is the exact cause is likely to have profound implications for cosmology and for particle physics.

Two recent collective reports, one by the US Dark Energy Task Force 
(DETF)\ \citep{detf},
convened by NASA, NSF and DOE, and another by the European ESA-ESO Working Group on Fundamental Cosmology
\citep{esa-eso},
have identified the most promising observational strategies to characterize DE properties in the near future. These reports concluded that the method based on measurements of Baryon Acoustic Oscillations (BAO) from galaxy redshift surveys is less likely to be limited by systematic uncertainties than other methods that are proposed.
It appears that, while recognizing the need for a combined strategy involving two or more independent techniques, BAO measurements can substantially contribute to increase the accuracy on the DE equation of state.

Baryon Acoustic Oscillations are produced by acoustic waves in the photon-baryon plasma generated by primordial perturbations \citep{eh98}. At recombination ($z\sim 1100$), the photons decouple from the baryons and start to free stream, whereas the pressure waves stall. As a result, baryons accumulate at a fixed distance from the original overdensity. This distance is equal to the sound horizon length at the decoupling time, $r_{BAO}$. The result is a peak in the mass correlation function at the corresponding scale. First detections of this excess were recently reported, at a significance of about three standard-deviations, both in spectroscopic~\citep{eisenstein,2dFb,hutsi} and photometric~\citep{nikhil,blake} galaxy redshift surveys.

The comoving BAO scale is accurately determined by CMB observations ($r_{BAO}=146.8 \pm 1.8$~Mpc  for a flat $\Lambda CDM$ Universe~\citep{Hinshaw08}), and constitutes a ``standard ruler'' of known physical length. The existence of this natural standard ruler, measurable at different redshifts, makes it possible to probe the expansion history of the
Universe, and thereby dark energy properties and the Universe geometry
(see, e.g., \citet{seo,Glazebrook} and references therein). This motivates the present efforts to measure BAO (e.g., \citet{adept}, \citet{des}, \citet{hetdex}, \citet{panstarrs}, \citet{space}, 
\citet{wfmos}, \citet{wigglez}).

Broad-band photometric galaxy surveys can measure the angular scale of
$r_{BAO}$ in several redshift shells, thereby determining
$d_{A}(z)/r_{BAO}$, where $d_{A}(z)$ is the angular distance to the
shell at redshift $z$. If galaxy redshifts can be determined precisely enough, the BAO scale can also be measured along the line of sight, providing a direct measurement of the instantaneous expansion rate, the Hubble parameter (or actually of $H(z)\, r_{BAO}$), at different redshifts. This quantity is more sensitive to the matter-energy contents of the universe compared to the integrated quantity $d_{A}(z)$. The direct determination of $H(z)$ 
distinguishes the BAO method from other methods. In addition, since systematic errors affect the radial and tangential measurements in different ways, the consistency between the measured values of $H(z)$ and $d_{A}(z)$ offers a test of the results.

As a rule of thumb, in order to get the same sensitivity to the dark-energy parameters, a galaxy redshift survey capable of exploiting the information along the line of sight needs to cover only $\sim$ 10\% of the volume covered by a comparable survey that detects the scale in the transverse direction only~\citep{briddle}. When covering a similar volume, precise enough redshift measurements can provide substantially tighter constraints on the DE parameters.

Large volumes have to be surveyed in order to reach the statistical accuracy needed to obtain relevant constraints on dark-energy parameters. Enough galaxies must be observed to reduce the shot noise well below the irreducible component due to sampling variance
(see section~\ref{sec:power}).
The usefulness of the correlation along the line of sight favors spectroscopic redshift surveys that obtain very accurate redshifts, but the need for a large volume favors photometric redshifts that can reach down to fainter galaxies.

The intrinsic comoving width of the peak in the mass correlation function is about 15~Mpc/h, due mostly to Silk damping~\citep{Silk}. This sets a requirement for the redshift error of order $\sigma(z)=0.003(1+z)$, corresponding to 15~Mpc/h along the line of sight at $z=0.5$. A much better precision will result in oversampling of the peak without a substantial improvement on its detection, while worse precision will, of course, result in the effective loss of the information in the radial modes \citep{seo}. Note also that in the presence of substantial redshift errors, the error distribution needs to be known and accurately corrected for when inferring the BAO scale.

It has usually been assumed in the literature that photometric redshifts are not precise enough to measure ``line-of-sight'' BAO \citep{seo,briddle}. While this is true for broad-band photometric surveys, here we examine how one can reach the required redshift precision with narrow-band photometry. We find (see section~\ref{sec:survey}) that redshifts of luminous red galaxies can be measured with a precision $\sigma(z)\sim 0.003(1+z)$ using a photometric system of $40$ filters of $\sim 100$~\AA, continuously covering the spectral range from $\sim 4000$ to $\sim 8000$~\AA , plus two additional broad-band filters similar to the $u$ and $z$ bands.

We describe the practical implementation of this idea, a photometric galaxy redshift survey called PAU (Physics of the Accelerating Universe). PAU will measure positions and redshifts for over 14 million luminous red galaxies over 8000 deg$^{2}$ in the sky, in the range $0.1<z<0.9$ (comprising a volume of
9~(Gpc/h)$^{3}$), and with an expected photometric redshift precision
$\sigma(z)\lesssim 0.003(1+z)$. This redshift precision makes it possible to measure radial BAO with minimal loss of information. The PAU survey can be carried out in a four
year observing program at a dedicated telescope with an effective {\it etendue} $\sim 20$~m$^2$deg$^2$.

The outline of the paper is as follows. Section~\ref{sec:req} discusses the scientific requirements for such a survey. An optimization of the survey parameters follows in section~\ref{sec:survey}, while section~\ref{sec:base} presents a possible baseline design for the instrument. The science capabilities of the survey are given in section~\ref{sec:science}. Finally, section~\ref{sec:summary} contains the summary and conclusions.

\section{Scientific Requirements}

\label{sec:req}

Given current priors on other cosmological parameters a measurement
of the expansion rate history $H(z)$ with percent precision will
translate into a measurement of the DE equation of state w
of few times this precision.
We will focus here on how well we can do this by measuring the BAO feature
at comoving size $r_{BAO}\simeq 100Mpc/h$ and use
it as a standard ruler in both the radial and tangential directions.

We first explore how an error in the BAO scale $r_{BAO}$ translates into
an error in the DE equation of state $w$. The relation is different
if we measure the scale in the radial or in the tangential (ie angular)
directions. We then move to show how the BAO scale can be measured
statistically using galaxy surveys. We start with a visual ilustration
of the problem and a brief presentation of two N-body simulations
that we used in order to study the main goal of this paper.
We then show how this scale can be measured using the
statistics of galaxy density fluctuations and relate the error in $r_{BAO}$
to the volume of the survey, given perfect distance indicators. 
We finish by considering what redshift precision is required to maintain
a given precision in the $r_{BAO}$ measurement and how this is limited
by different systematic effects.
Note that we focus here in showing the implication of the photo-z accuracy 
in the measurements of the monopole, which combines the radial and perpendicular 
information. These two components can be separated by considering the anisotropic 
correlation function
(see Okumara et al.~2007, Padmanabhan \& White 2008, Cabre \& Gaztanaga 2008, and
Figs.~17--18 in Gaztanaga, Cabre \& Hui 2008, 
which show how the anisotropic correlation 
function changes for PAU-like photo-z precisions).

We note that these are just rough estimates
to show the viability of this approach. In a real survey, there might be
other sources of systematic errors that have not been taken into account in 
detail here. However, experience indicate that the actual data themselves can 
be used to study and minimize those sources of errors.

In what follows we assume a flat FRW cosmology, with cosmological
parameters compatible with WMAP data~\citep{Hinshaw08}.

\subsection{BAO scale and DE equation of state}

 In a galaxy survey we measure distances in terms of angles and redshifts.
These observed quantities can be related to known distances, such
as $r_{BAO}$, using the FRW metric.
The differential radial (comoving)
distance is inversely proportional to the expansion rate
$H(z) \equiv \dot{a}/a$:
\begin{equation}
dr(z)=\frac{c}{{H(z)}}~dz\ ,
\end{equation}
while the angular diameter distance is proportional to the integral
of $dr(z)$:
\begin{equation}
d_{A}(z)=\frac{c}{{1+z}}\int_{0}^{z}~\frac{dz'}{{H(z')}}
\end{equation}
for a flat universe.
In particular, measurements of the charateristic size of the
BAO feature in the radial ($\delta z_{BAO}$)
and tangential ($\delta \theta_{BAO}$) direction relate to the known comoving
BAO scale $r_{BAO}$ as:
\begin{eqnarray}
\delta z_{BAO} & = & r_{BAO}~\frac{H(z)}{{c}}\\
\delta \theta_{BAO} & = & \frac{r_{BAO}}{{d_{A}(z)}(1+z)}
\end{eqnarray}
We therefore have, neglecting for a moment the uncertainty on the determination of $r_{BAO}$
by CMB observations, that a relative error in the measured size
of the BAO feature corresponds to a relative error in either $d_A(z)$ or $H(z)$:
\begin{eqnarray}
\Delta^T_{BAO} \equiv \frac{\sigma(\delta\theta_{BAO})}{\delta\theta_{BAO}}
= \frac{\Delta d_A }{d_A} \nonumber \\
\Delta^L_{BAO} \equiv \frac{\sigma(\delta z_{BAO})}{\delta z_{BAO}} =
\frac{\Delta H}{H}
\label{deltaBAO}
\end{eqnarray}

\begin{figure}
\epsscale{1}
\plotone{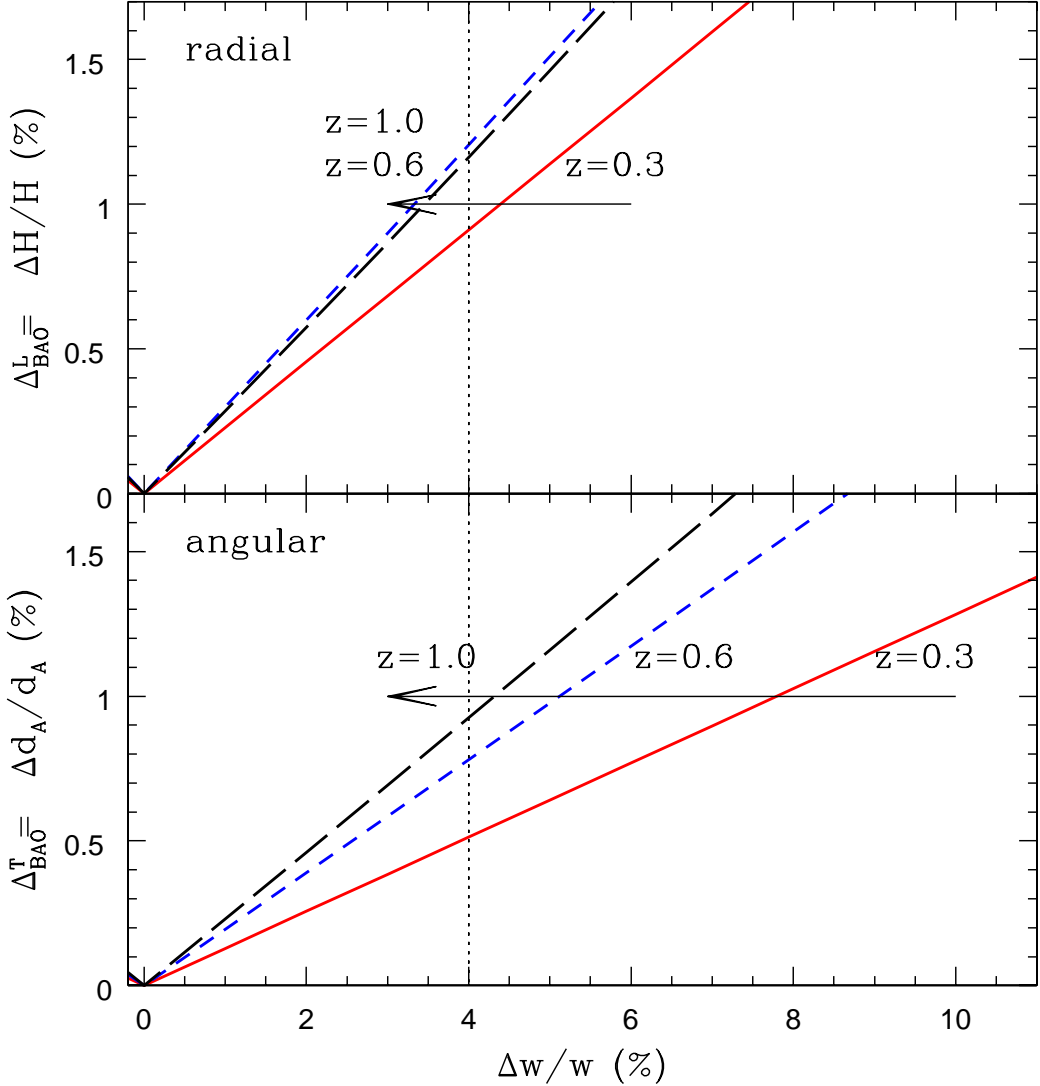}
\figcaption[walfa]{Relation between the change in the
dark energy equation of state parameter $w$, shown in the x-axis,
and its effect in the apparent measured BAO scale,
$\Delta_{BAO}$, shown in the y-axis.
Changes are shown in percent,  relative to the w=-1 case.
Top panel corresponds
to the radial distance: $\Delta^L_{BAO}=\Delta H(z)/H(z)$. The bottom panel
shows the angular diameter distance: $\Delta^T_{BAO}=\Delta d_{A}(z)/d_{A}(z)$.
The different lines correspond to $z=0.3$ (continuous),
$z=0.6$ (short dashed) and $z=1$ (long dashed).
In all cases $\Omega_{m}=0.25$ and flat universe are assumed. All other cosmological
parameters are kept fixed.
\label{fig:walfa}}
\end{figure}

For this argument, we will write the expansion rate as:
\begin{equation}
H^{2}(z)/H_{0}^{2}=\Omega_{m}\;(1+z)^{3}+\left(1-\Omega_{m}\right)\;(1+z)^{3(1+w)}
\end{equation}
which corresponds to a flat universe ($\Omega_{m}+\Omega_{DE}=1$)
with a constant equation of state $w\equiv p/\rho$, with $p$ the
pressure, and $\rho$ the density of the dark energy. Figure~\ref{fig:walfa}
shows the relative change in Eq.(\ref{deltaBAO}) (in percent)
as a function of the relative changes in $w$
with respect to $w=-1$  for different redshifts. We show the cases
for $z = 0.3-1.0$ which will be relevant for our study. As can be seen in
the figure, a $1\%$ error in $\Delta_{BAO}$ (our goal)
at $z \simeq 1.0$ results in a $~\simeq 4\%$
uncertainty in $w$, but the precision varies with redshift.
A constant 1\% error in the angular distance quickly degrades the $w$ precision
with decreasing redshift,
from $4.5\%$ at $z=1.0$ to $8.0\%$ at $z=0.3$, while
the radial distance achieves a more uniform precision
in $w$, $3.5-4.5\%$ in the whole redshift range.

This illustrates the advantage of having a good radial
measurement. Angular distances provide a good geometrical test, while
radial distances tell us directly the instantaneous expansion rate.
In addition, comparing relative sizes of the BAO feature when measured
parallel and perpendicular to the line of sight will provide us with
a consistency test.

\subsection{BAO scale in N-body simulations}

To support some of the main claims of this paper
we have used large N-body dark matter (DM) simulations, 
using the MICE collaboration\footnote{http://www.ice.cat/mice} set-up.
In particular, we have computed non-linear DM clustering statistics in terms of 
the 2-point correlation function, $\xi(r)$, and its Fourier transform, 
the power spectrum, $P(k)$, and we have assessed the impact of real world systematic effects 
on the BAO mesurements based on these standard estimators.

MICE simulations have been run 
using the Gadget-2 code \citep{Springel05} on the MareNostrum supercomputer
at BSC \footnote{Barcelona Supercomputer Center, http://www.bsc.es},
with a modification to produce outputs in the light-cone
\citep{OnionUniverse}.  We focus here on two simulations, shown in Table \ref{tableN},
corresponding to a flat concordance $\Lambda$CDM model with $\Omega_m=0.25$,
$\Omega_\Lambda =0.75$, $\Omega_b=0.044$, $n_s=0.95$, $\sigma_8=0.8$
and $h=0.7$. 
Halos were obtained from the $z=0.5$ comoving output using the Friends-of-Friends (FoF) 
algorithm with linking length $0.164$. 
The larger simulation has a dynamic range close to five orders of magnitude.

\begin{figure}
\epsscale{1}
\plotone{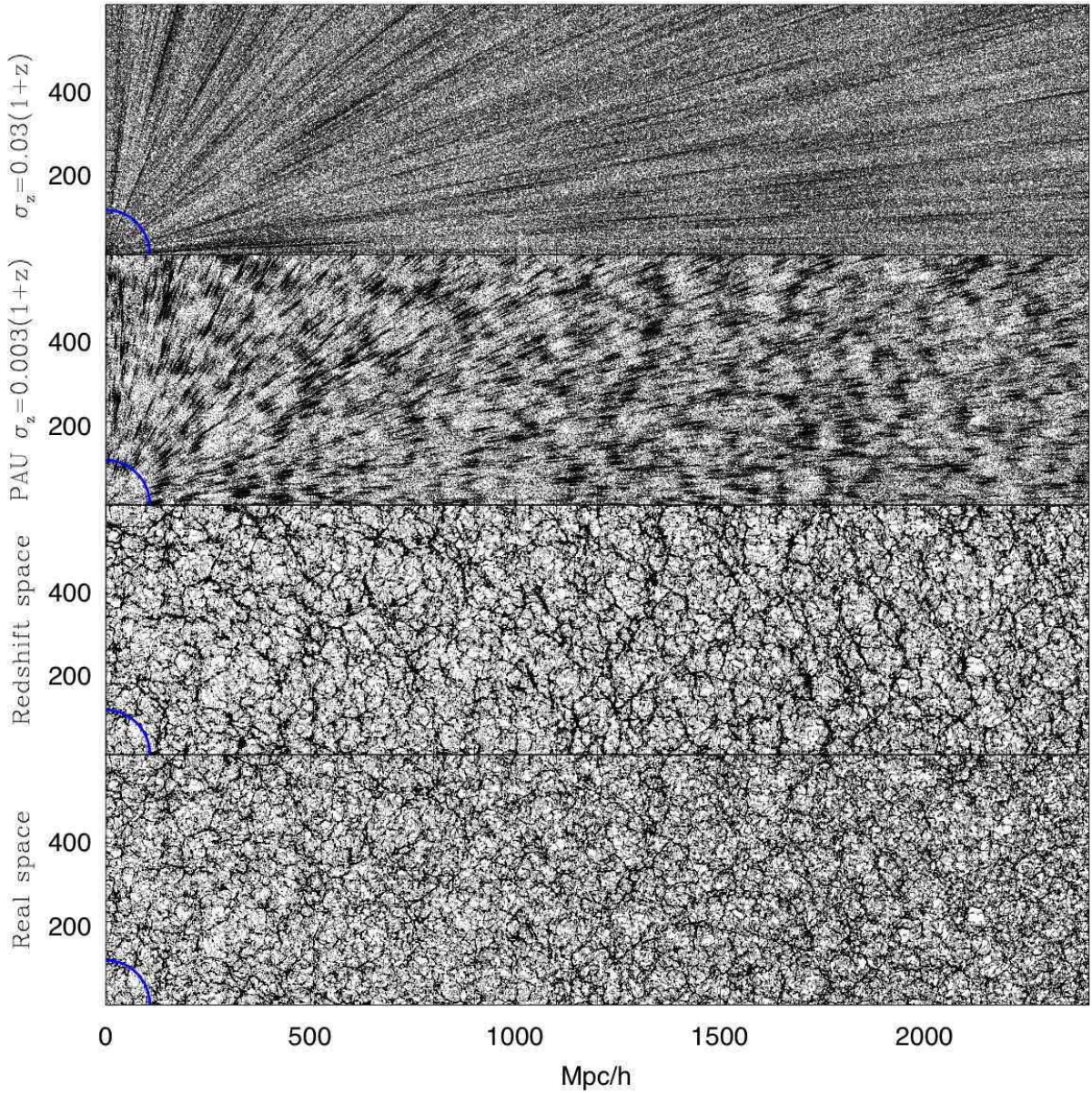}
\figcaption[slice]{Sytematic effects in the lightcone:  
panels show a 1 Mpc/h thick section of the lightcone
distribution in MICE3072 in comoving coordinates.
The two bottom panels corresponds to the
actual dark matter distribution in the simulation in real (bottom) and
redshift space (upper panels). The top two panels also include 
a (Gaussian distributed) photo-z error distortion
of $\sigma_z = 0.003(1+z)$, as expected from PAU galaxies, and
an order-of-magnitude worse case, $\sigma_z = 0.03(1+z)$.
The BAO scale is shown by a section of circle with radius $~100$ Mpc/h around the observer. 
\label{slice}}
\end{figure}

Figure~\ref{slice}
shows a thin slice of the light-cone built from the MICE3072 simulation.
We build the light-cone placing the observer at the origin, so that cosmic time (redshift)
corresponds to the radial direction, which expands from $z=0$
to $z=1$ (corresponding to a comoving radius of 2400 Mpc/h for
our cosmology).  The bottom panel corresponds to the
true dark matter distribution in real space.
The next panel up shows the redshift space distribution,
where the radial positions are distorted due to peculiar (gravitationally induced) 
motions away from the Hubble flow.
In order to model this distortion, we add the radial component of
the peculiar velocity $v_r$ of each particle to its (real space) position:
$s = r + f v_r (1+z)/H(z)$, where $f\sim 1$ for the assumed $\Lambda$CDM cosmology at $z\sim 0.5$.
We note that, in this image, distortions can only be seen
whenever they are much larger than the pixel size $\Delta r_{pix} \simeq 3 \rm{Mpc/h}$ 
or, in velocity units, $\Delta v_{pix} \simeq 300 \rm{km/s}$.
This implies that the so called Fingers of God effect (see explanation below) cannot be detected 
because it arises from random peculiar velocities of order $\Delta v_{pix}$. Instead, the 
Kaiser effect \citep{Kaiser} due to a large-scale coherent infall is visible as an 
enhancement of the filamentary structures perpendicular to the line of sight. 

The two top panels include in addition a radial distortion
due to photo-z errors which we assumed to be Gaussian distributed. Thus,  
they can be modeled by randomly displacing the particles along the line of sight
according to the probability distribution, 
\begin{equation}
f(\delta r_{z})\sim\exp\left[-(1/2)\left(\delta r_z/\Delta_z\right)^{2}\right],
\end{equation}
with the smoothing scale $\Delta_z$ and the photo-z error
$\sigma_{z}$ related through the Hubble parameter $H(z)$ as $\Delta_z=\sigma_z (1+z)c/H(z)$.
Third panel up assumes $\sigma_z = 0.003(1+z)$, which is roughly the photo-z error expected for PAU
galaxies, while the top panel corresponds to an order-of-magnitude worse case, $\sigma_z = 0.03(1+z)$.  

Overall this figure illustrates how the image is degraded both by peculiar velocities and
redshift errors. Comparing the middle panels, 
it is evident that redshift errors produce, on average, 
much stronger distortions than peculiar velocities.

\subsection{Redshift errors}
\label{sec:redshifterrors}
 
We now turn to a more quantitative estimate of the minimum radial resolution required to detect the BAO scale in 3D. 

The BAO signature appears in the two-point correlation of particles as a single bump
at a scale $r_{BAO}\simeq 100\, h^{-1}\,{\rm Mpc}$ with an intrinsic
width $\Delta r_{BAO}\simeq 10\, h^{-1}\,{\rm Mpc}$ and relative amplitude
of about a factor two with respect to a non-BAO model with the same broad-band
shape. Therefore, a simple approach would be to look at the degradation of this
peak as we increase the redshift error.

To be more realistic, we have studied the clustering of halos
(selected with a FoF algorithm with linking length of 0.164) 
since they are more closely related to the observed galaxies and clusters of galaxies 
than DM particles. 
In particular, Luminous Red Galaxies
(LRGs) are thought to populate large DM halos and closely trace the
halo distribution. 

At large scales the halo and matter density fluctuations are related by a linear bias factor 
($\delta_{halo} = b\,\delta_{matter}$), which translates into a $b^2$ scaling
for the 2-point funcion. Halos with large mass cut in Table \ref{tableN} have
$b\simeq 3$, which was chosen to magnify possible non-linear effects.
To estimate errors (eg.~in Fig.~\ref{fig:Pk}), we use instead a lower mass cut which
corresponds to $b \simeq 2$ and better matches the clustering of LRGs that
have already been used to measure BAO \citep[e.g.][]{eisenstein,Tegmark06}.

\begin{figure}
\epsscale{1}
\plottwo{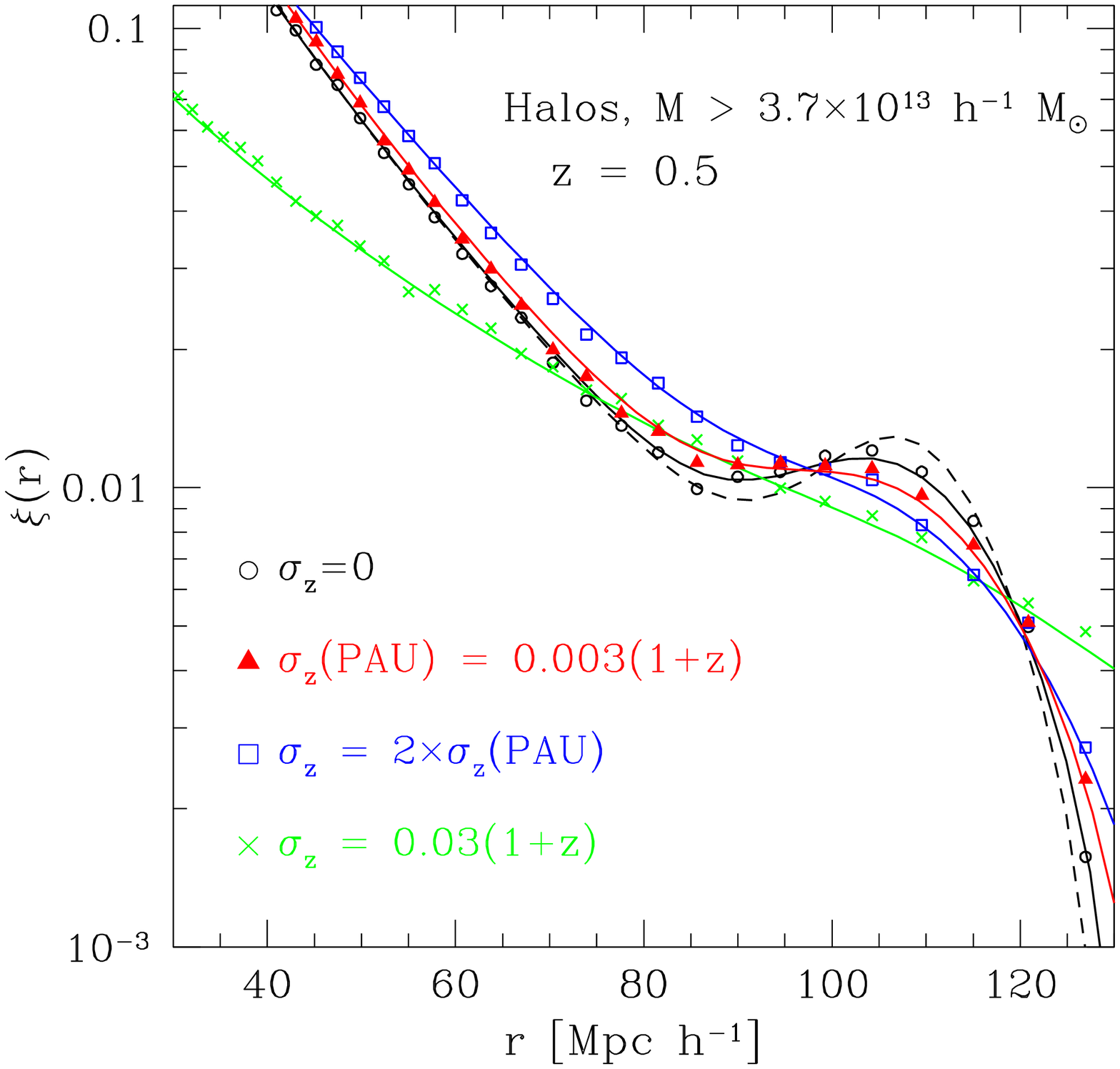}{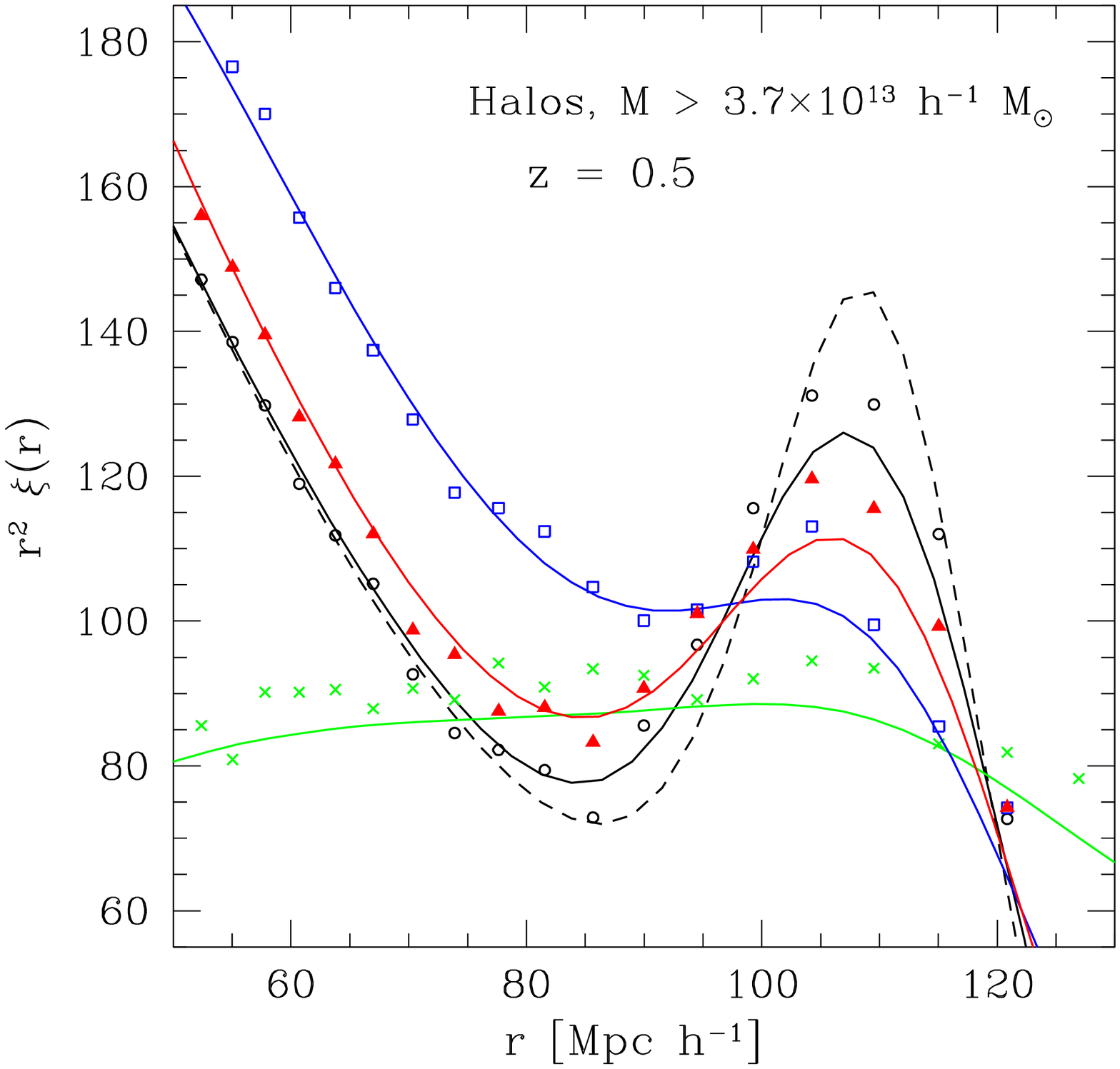}
\figcaption[xi2]{
Smearing of the BAO signature due to photometric redshift errors. The circles
denote the two-point correlation function from over a million halos
with mass $M>3.7\times10^{13}\, h^{-1}\, M_{\odot}$ (assumed to
host LRGs) measured in a MICE simulation of
$27\, h^{-3}\,{\rm Gpc}^{3}$ volume.
The dashed line is the linear correlation function scaled with the
linear halo bias ($b=3$), while the black solid line corresponds
to the nonlinear prediction given by RPT \citep{RPTbao}.
Their difference shows the degradation coming solely
from nonlinear clustering. In addition, the triangle (red), 
square (blue) and cross (green) symbols show the measured correlation 
function after a Gaussian error degradation
in the line-of-sight position of the halos is introduced
($\sigma_{z}/(1+z)=0.003,0.007$
and $0.03$, respectively). The corresponding solid lines are the analytical
predictions derived from Eq.~(\ref{eqPphotoz}). The right panel shows a
zoom over the peak region scaled as $r^2\xi(r)$. Clearly, the signal-to-noise
in the BAO feature reduces with photo-z error and starts to totally disappear
above the PAU threshold of $0.003$, which
roughly corresponds to the intrinsic width of the BAO feature due to
Silk damping.
\label{fig:xi2}}
\end{figure}

Figure~\ref{fig:xi2} shows the two-point correlation function (black empty circles) traced
by all halos in MICE3072 at $z=0.5$ (see Table\ref{tableN}). The dashed line corresponds to the linear correlation and the black solid line to the nonlinear model of Renormalized
Perturbation Theory (RPT) \citep{RPTbao},
both biased with $b=3$. As illustrated by
Fig.~\ref{fig:xi2} (solid black line), 
the simple scale independent linear bias scheme works reasonably well, but this assumption certainly needs to be tested more accurately.

We then estimated the impact of photo-z errors using Gaussian distortions in the radial direction as described in the previous section.
The corresponding correlation function of this smeared distribution of halos
is shown in Fig.~\ref{fig:xi2} for different values of the photo-z error,
$\sigma_{z}/(1+z)=0.003$ (as expected for PAU) $0.006, 0.03$, with
red, blue and green symbols respectively.

In turn, this smearing can be modeled in Fourier Space by damping
the power spectrum along the line-of-sight as
\begin{equation}
P_z(k,\mu)=b^2 P_{nl}(k)\exp\left[-k^{2}\Delta_z^{2}\mu^2\right],\label{eqPphotoz}
\end{equation}
where $P_{nl}$ is the nonlinear power spectrum from RPT, the linear bias is $b=3$, and $\mu$ is the cosine of the angle with the line of sight. The red, blue and green solid lines
in Fig.~\ref{fig:xi2} correspond to the angle averaged Fourier transform
of Eq.~(\ref{eqPphotoz}) for $\sigma_{z}=0.003$, $0.006$ and $0.03$
respectively. 

In summary, Fig. ~\ref{fig:xi2} illustrates that one can basically
recover the right BAO shape once the error is better than about $0.003(1+z)$.
Larger errors erase the BAO bump and will result in the loss of cosmological
information. The change can be roughly quantified by the ratio between
the amplitude of the BAO peak (at $r \simeq 108$ Mpc/h) and the amplitude in
the valley (at $r \simeq 85$ Mpc/h). For $r^2 \xi(r)$ in the right panel
of Figure ~\ref{fig:xi2},  this ratio is about $1.8$ in real space
with no photometric errors and decreases smoothly to $1.5$ as we increase
the error towards $0.003(1+z)$.
For larger errors, this ratio decreases more rapidly an gets all the
way to unity for $0.006(1+z)$, as shown in the figure.
This makes sense because $0.003(1+z)$ corresponds
to a comoving scale of about $15\, h^{-1}\,{\rm Mpc}$ at $z=0.5$, which matches the
intrinsic width (Silk damping) of the BAO peak. Although this is
just a rough estimate, it is all we need as a starting point
for our considerations below, and it agrees with other considerations
based on counting the number of modes in 2D and 3D surveys and the
work of \citet{Glazebrook,seo}.

The clustering analysis presented above is in real rather than redshift space.
Redshift space distortions can
be modeled as a combination of two separate effects:
coherent and random peculiar velocities.
The first term is the so-called Kaiser effect \citep{Kaiser},
which increases the amplitude
of clustering at large scales by a factor $\sim (1+\beta \mu^2)^2$ (where $\beta=\Omega_m^{0.6}/b$, $b$ is the bias and $\mu$ the cosine of the angle with the line of sight). Our analysis allows for such an effect by incorporating a larger effective bias in the correlation function monopole (e.g. $b=3$ as in Fig.~\ref{fig:xi2}). 
The  effect of random velocities can be modeled as a Gaussian
damping, very similar to photo-z errors in Eq.(\ref{eqPphotoz}) but where
$\sigma_z^2$ is replaced by $\sigma_p^2/2$, where $\sigma_p$ is
the one dimensional galaxy pairwise velocity
dispersion (the factor of 1/2 is because a velocity difference has twice the variance
of a single velocity).  The net effect is that the density field is convolved with a one
dimensional random field with a net dispersion
$\sigma^2=\sigma_p^2/2+\sigma_z^2$.
The typical value of $\sigma_p$ in our DM simulation is smaller than the photo-z errors
considered here. This is also the case in regions of high density, populated by LRG galaxies,
and where $\sigma_p$ could be larger, e.g, $\sigma_p \sim 400 \rm{km/s}/c \sim 0.0013$ for 
$r > 5 \rm{Mpc/h}$ \citep{Ross}. 
This is clearly seen by comparing the
two middle panels of Fig.\ref{slice} where it is evident that even in
regions of high density the photo-z distortions are larger than the
redshift space distortions.
Thus effectively $\sigma_z>\sigma_p$ and we can
consider this effect subdominant in our considerations. 

We note here that there is a fundamental limitation as to how much
one can improve the BAO measurement by reducing the photo-z error. The photo-z
error $\sigma_z \simeq 0.003(1+z)$
proposed by the PAU survey is close to optimal.
Redshift space distortions and non-linear effects can produce distortions that are
comparable to this value, depending on what is the (biased) tracer
that is used to measure BAO.

\subsection{Estimating the BAO scale}
\label{sec:power} 

Armed with the conclusions from the previous discussion about photo-z errors and other nonlinear effects 
(clustering, bias and redshift distortions), we are now in a position to give an estimate of the 
expected 1-$\sigma$ error determination from a survey with the characteristics of PAU. 
To this end we will employ the 2-point statistics in Fourier Space (i.e. the power spectrum $P(k)$).

To estimate the error in the measurement of the power spectrum we will resort to the commonly used expression:
\begin{equation}
\sigma_{P}\equiv\frac{\Delta P(k)}{{P(k)}}\simeq\sqrt{\frac{2}{{N_{m}(k)}}}\left(1+\frac{1}{\bar{n}P(k)}\right),
\label{eq:deltaP}
\end{equation}
which can be derived from~\citet{FKP94} (see, for instance, 
   Martinez \& Saar 2002, \S 8.2),
where $N_{m}(k)$ is the number of Fourier modes present in a spherical
shell extending from $k$ to $k+\Delta k$. In terms of the survey
volume $V$, we have $N_{m}(k)=V(4\pi k^{2}\Delta k)/(2\pi)^{3}$.
The first term in Eq.~(\ref{eq:deltaP}) corresponds to the sampling error and is independent of
redshift. The second term corresponds to Poisson shot-noise, and $\bar{n}$
denotes the number density of observed galaxies in the survey. This
formula is exact when the probability density function of spectral
amplitudes is Gaussian and a very good approximation when the shot-noise 
term is negligible~\citep[see][]{ABFL07}. 

For $k\le0.12\,h^{-1}\,{\rm Mpc}$ and for halos that host LRGs 
we expect that $P(k)>2\times 10^4\,h^{-3}\,{\rm Mpc}^3$ ($b\simeq 2$) at $z\sim 0$. This agrees well with actual measurements of $P(k)$ for LRGs in the SDSS catalogue (see e.g. Fig.~4 in \citet{Tegmark06}). As we will show below, the PAU number density of LRGs is expected to be $\bar{n}> 0.001\,h^3\,{\rm Mpc}^{-3}$ for $z<0.9$ (see Fig.\ref{fig:nv} below), which implies that the Poisson shot-noise contribution to the error in Eq.(\ref{eq:deltaP}) is smaller than $8\%$ at $z\sim 0.5$, even taking into account the degradation from a photo-z error of $\sigma_z = 0.003$ as discussed before. 

\begin{figure}
\epsscale{1}
\plottwo{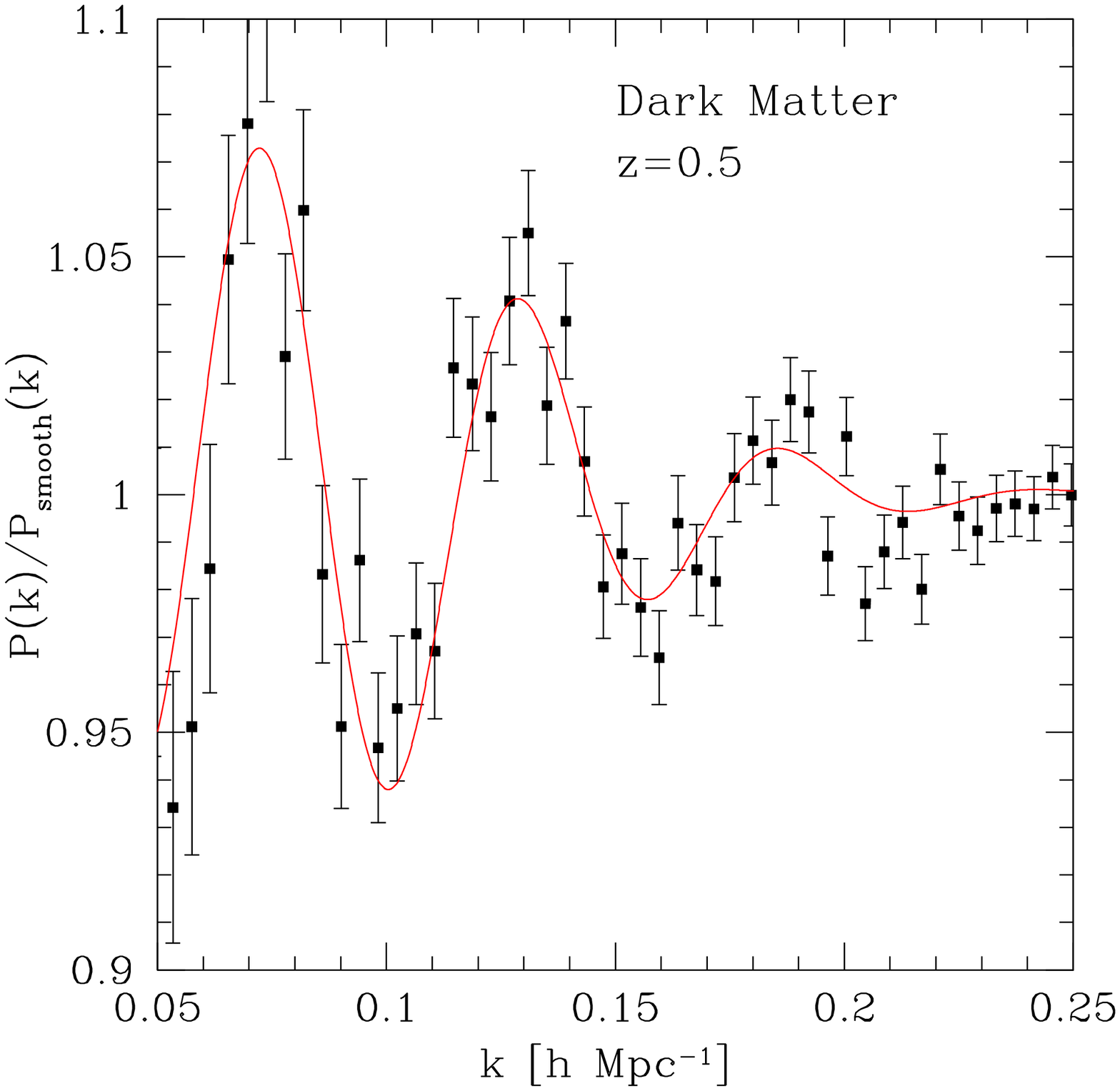}{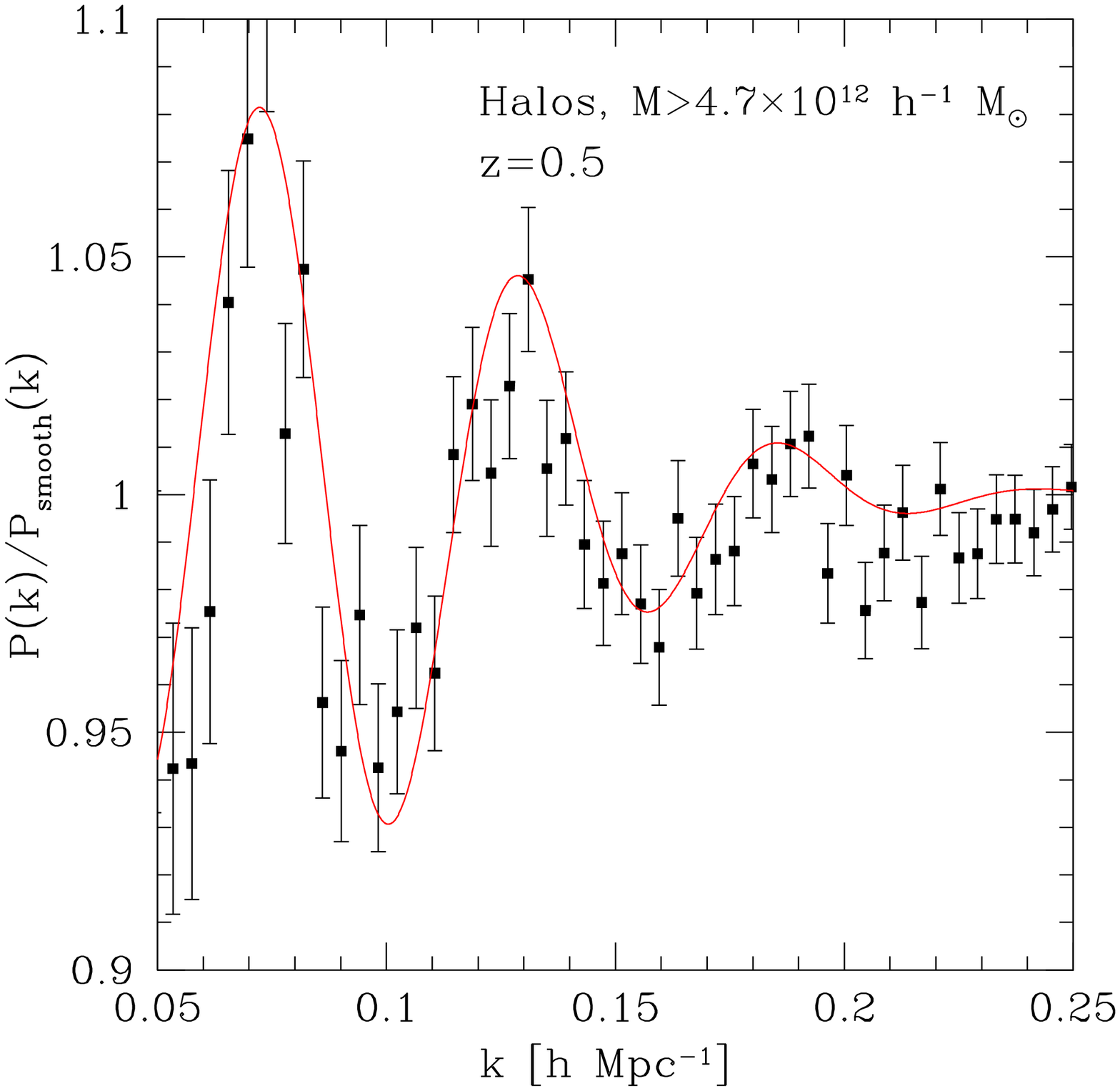}
\figcaption[Pk]{Ratio of the power spectrum measured at $z=0.5$ in MICE1536
to a smoothed version of the same spectrum
for dark matter (left panel) and halos of mass $M\ge4.7\times10^{12}\, h^{-1}\, M_{\odot}$ (right panel). These large halos are expected to host LRGs.
The reference spectrum was obtained by rebinning the
original one in bins of $\Delta k=0.055\, h\,{\rm Mpc}^{-1}$ in order
to wash out the BAO signature but keeping the broad band shape of the nonlinear spectrum.
Error bars were obtained using the
approximation in Eq.~(\ref{eq:deltaP}). The red solid line
corresponds to the parametric fit given by Eq.~(\ref{parametric})
with $A / r_{BAO}=0.016 $ for dark matter and 
$A / r_{BAO}=0.017$ for halos 
($r_{BAO}=108.6\,\mathrm{Mpc/h}$ for our reference cosmology). 
This figure illustrates that for both
dark matter and halos, one can approximately model BAO in the $P(k)$ 
with Eq.~(\ref{parametric}). This conclusion also applies in redshift
space and for different galaxy populations \citep{ABFL07}. \label{fig:Pk}}
\end{figure}

The bump in the spatial correlation function translates into the power spectrum as a series of damped oscillations of a few percent in relative amplitude. This is shown in Fig.~\ref{fig:Pk}  that contains the power spectrum of DM (left panel) and halos of mass $M>4.7\times10^{12}\, h^{-1}\, M_{\odot}$ (right panel) measured in the MICE1536 simulation. In both panels the measured spectra have been divided by a smoothed one with the same broad band power obtained from the data themselves~\citep[][]{2dFb,ABFL07}.  The solid red lines in  Fig.~\ref{fig:Pk} show that this ratio can be roughly modeled as,
\begin{equation}
\hat{P}(k)\simeq1+Ak\exp[-(k/0.1h{\rm Mpc}^{-1})^{2}]\sin{(r_{BAO}k)},\ \
\label{parametric}
\end{equation}
which illustrates how $P(k)$ depends on the BAO scale.

Moreover, the discussion in Sec.~\ref{sec:redshifterrors} that led to Fig.~\ref{fig:xi2} validates to a good extent that photo-z, clustering, bias and redshift distortions can be modelled in the power spectrum monopole as the angle average of,
\begin{equation}
P(k,\mu)= b^2 P_{nl}(k) (1+\beta \mu^2)^2 \exp\left[-k^2 \Delta^2 \mu^2\right],
\label{fullPk}
\end{equation}
where $\Delta= \sigma (1+z)/ H(z)$ and $\sigma=\sqrt{\sigma^2_z+\sigma^2_p/2}$. Therefore, except from nonlinear clustering, which is stronger than the intrinsic Silk damping of BAO, and is responsible for the exponential damping in Eq.~(\ref{parametric}), the remaining effects are multiplicative contributions to the measured $P(k)$, and they factor out when constructing $\hat{P}$. 
This means that the above described systematic effects do not affect the BAO signal 
in the spherically averaged $P(k)$, although they do increase the associated errors as 
we will discuss below.
Thus Eq.~(\ref{parametric}) allows to compute how the measured $\hat{P}$ 
varies with the BAO scale, 
\begin{equation}
d\hat{P}/dr_{BAO}\simeq Ak^{2}\exp[-(k/0.1h{\rm Mpc}^{-1})^{2}]\cos{(r_{BAO}k)}
\end{equation}

Such variation in $\hat{P}$ produces a shift in the $\chi^{2}$ fitting to measurements
of $\hat{P}(k_{i})$ given by,
\begin{eqnarray*}
\Delta\chi^{2}\simeq\sum_{i}\frac{\Delta\hat{P}^{2}(k_{i})}{{\sigma_{P}^{2}(k_{i})}}\simeq\Delta_{BAO}^{2}\left(\frac{A}{{r_{BAO}}}\right)^{2}\!\left(\frac{V}{{r_{BAO}^{3}}}\right)\! I^{2}[m]\end{eqnarray*}
where,
\begin{eqnarray}
I^{2}[m]=\frac{1}{{(2\pi)^{2}}}\int_{0}^{2\pi m}\!\!\! \frac{x^{6}\exp[-2(x/10.86)^{2}]\cos^{2}(x)}{(1+1/ \bar{n} P)^2},\ \ \
\label{eqIm}
\end{eqnarray}
$m$ is the number of BAO oscillations included in the fit ($k_{max}=2\pi m / r_{BAO}$) and
$\Delta_{BAO}\equiv\Delta r_{BAO}/r_{BAO}$. The shot-noise term includes the full power spectrum given by Eq.~(\ref{fullPk}) and accounts, in particular, for photo-z errors. In deriving Eq.~(\ref{eqIm}) we have explicitly used that for our reference cosmology $r_{BAO}=108.6\, h^{-1}{\rm Mpc}$.

A 1-sigma determination of $\Delta_{BAO}$ alone corresponds to $\Delta\chi^{2}=1$,
so that,
\begin{equation}
\Delta_{BAO}|_{\Delta\chi^{2}=1}=
\left(\frac{r_{BAO}^{3}}{{V}}\right)^{1/2}\frac{1}{I[m](A/r_{BAO})}.
\label{estimate}\end{equation}
From Fig.~\ref{fig:Pk} we find that $A/r_{BAO}\sim0.02$ fits well
both halos and dark matter clustering. We then assume $m\simeq2.5$
which corresponds to the range $0<k<0.14~h~{\rm Mpc}^{-1}$ (including two BAO peaks) and obtain
from Eq.~(\ref{estimate}),
\begin{equation}
\Delta_{BAO}\simeq 0.33\left(\frac{r_{BAO}^{3}}{{V}}\right)^{1/2}
\simeq 0.33\% \sqrt{13\, h^{-3}\,{\rm Gpc}^{3}/V}
\end{equation}
when we neglect completely the shot-noise term in Eq.~(\ref{fullPk}). Including the shot-noise but no photo-z degradation yields a prefactor of $0.35\%$ (for $\sigma_p = 400 \rm{km/s}$). 
If we also add a photo-z error of $\sigma_z=0.003$ (as PAU) we find $0.36\%$. If instead, we add a photo-z error of $\sigma_z=0.03$, the amplitude rises to $2.2\%$. 
In other words, we expect PAU to yield a measurement of BAO with only a $10\%$ degradation with respect to an ideal survey, whereas a survey with an order-of-magnitude larger photo-z is expected to produce a factor $\sim 6.5$ worse measurement.
According to this, in a 3D analysis, the relative error in the BAO scale
is just approximately equal to the inverse of the square root of
the number of independent
regions of size $r_{BAO}^{3}$ that are sampled by our survey, and 
it is quite robust in front of close to optimal (PAU-like) photo-z error and nonlinear effects.
For $V \simeq 10 h^{-3}\,{\rm Gpc}^{3}$  we get about $\Delta_{BAO} \simeq 0.5\%$.
The above estimate is in good agreement with Table~2 in~\citet{ABFL07}
and with the analysis in~\citet{Glazebrook} and \citet{seo}.

If we limit ourselves to optical surveys of LRGs,
we have $z \lesssim 1$. To get to $V\simeq 10\, h^{-3}\,{\rm Gpc}^{3}$
we will have to map of the order of $8000$~sq.~deg. There
are roughly two million LRGs with luminosity $L$ above the characteristic 
galaxy luminosity $L^\star$ in 1000 sq.~deg.~at $z<0.9$ with
magnitude $I_{AB}<22.5$~\citep{Brown2007}.
However, not all galaxies in a given
volume need to be measured as long as $\bar{n}P>3$, so that shot
noise is sub-dominant in Eq.~(\ref{eq:deltaP}). We will show below
that it is in fact possible to get to $\bar{n}P \gtrsim 10$ 
with the subsample of PAU LRGs that have good quality redshifts.

\section{Survey Simulations}\label{sec:survey} 

\begin{figure}
\epsscale{1}
\plotone{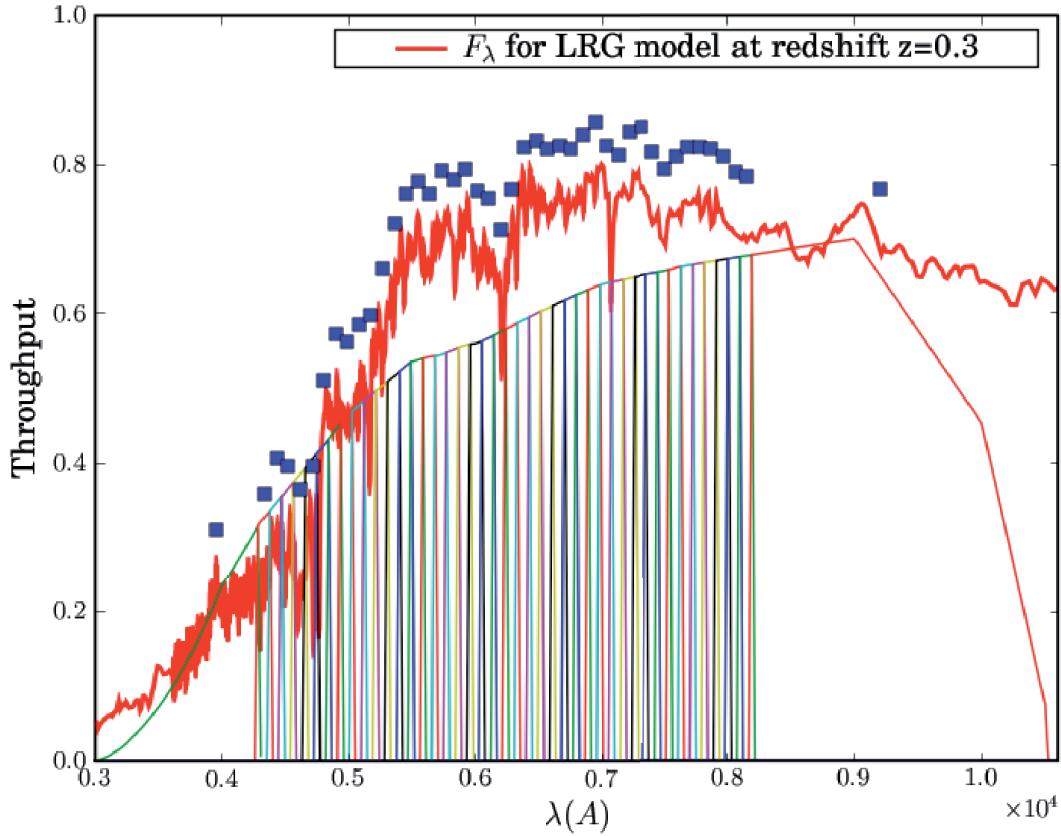}
\figcaption{An example of a filter system similar to the one which will be used
by the PAU survey. We have included the redshifted spectrum of an early type
galaxy at z=0.2 from the Bruzual and Charlot library to illustrate
how the sharp 4000~\AA\ break (which here falls at 4800~\AA) 
is basically bracketed by only two filters. Note that the filters
are spaced 93~\AA\ but have FWHM widths of 118~\AA\ 
due to the wavelength extent of their wings. The blue squares represent 
the flux which would be observed through the filters. Note that many 
spectral features apart from the 4000~\AA\ break are resolved by such a 
filter system. 
\label{fig:filters}}
\end{figure}

  The main distinctive feature of our survey is the use of photometric
information to achieve the highly accurate redshift measurements
needed to characterize the line-of-sight BAO signature.  Since such an
observational program has not been attempted before, we need to prove,
at least conceptually, that it is possible to achieve precisions of
$\sigma_z/(1+z)\simeq0.003$ with photometric data. According to
section~\ref{sec:req} and particularly Eq.~(\ref{eq:deltaP}), we only
need to reach such a precision for a galaxy population tracer with a
space number density which satisfies $\bar{n}P(k) \gtrsim 3$. Ideally
we would like galaxies that are luminous so we can observe them up to
high redshift and that present a spectral energy distribution with
distinctive features to achieve accurate photometric redshifts. The
most luminous of the early type galaxies (LRGs or Luminous Red
Galaxies) constitute such a population. Their space number density is
high enough. They are highly biased and they feature a prominent
4000~\AA\ break in their spectrum which, together with other spectral
features, makes possible a precise estimation of their redshifts using
photometric measurements (Fig.~\ref{fig:filters}). As a matter of fact,
bright early type galaxies were the subject of the first attempt to
estimate photometric redshifts in the seminal paper of \citet{1962IAUS...15..390B}.

\citet{Hickson1994} was the first to propose using intermediate band
filters as a viable alternative to traditional spectroscopy. The
COMBO-17 survey \citep{Wolf2001,Wolf2003} put this
idea into practice, using a combination of traditional broad band and
medium band filters. COMBO-17 reaches an accuracy of $\sigma_z\sim
0.02(1+z)$ for the general galaxy population~\citep{Hildebrandt2008},
and has reached a scatter of $\sigma_z\sim 0.0063$ for the bright
ellipticals in the Abell 901/902 superclusters.  Taking into account
the velocity dispersion of the cluster the authors infer an intrinsic
photometric accuracy close to $0.004(1+z)$ \citep{Wolf2003}.

\citet{Benitez2008} has shown that the most effective way of reaching
high photo-z precisions is using a system of constant-width,
contiguous, non-overlapping filters. The ALHAMBRA
survey~\citep{ALHAMBRA,Moles2008} has implemented such a filter
system, and preliminary results for that survey show that it is
possible to get close to $0.01(1+z)$ photo-z accuracy for the general
galaxy population.  LRGs usually have higher photo-z precisions than
the rest of the galaxies, (as it happens with the COMBO-17 data or the
SDSS LRGs \citep{Oyaizu2008, D'Abrusco}), and it is expected that
LRGs in the ALHAMBRA survey will have photo-z
errors substantially below $\sigma_z\sim 0.01(1+z)$.

  In view of these results, it seems reasonable to suggest that a
precision a few times smaller that $0.01(1+z)$ can be reached for the
LRG population with filters that are three times narrower than those
of ALHAMBRA, and about 2-3 times narrower than the medium band filters
in COMBO-17 (which do not fill the optical range contiguously). In
what follows we try to demonstrate that the observing program required
by PAU is feasible,
and can deliver redshift values with $\sigma_z/(1+z) \simeq 0.003$ for LRGs.

 In order to qualitatively understand the relationship between
measurement uncertainties and the accuracy of the redshift estimation
we can use a toy, step-like spectrum, flat in wavelength except for a
jump by a factor $D$ in the amplitude at $4000$~{\AA}: i.e. we assume that
the spectrum has a flux of $F$ redwards of the break and $F/D$
bluewards of it. This roughly approximates a low resolution version of
an LRG~\citep{EisensteinLRG}. If we use a set of constant
width, contiguous filters of width $\Delta\lambda$, the flux in the
filter that spans the break will be equal to
\begin{equation}
f=\alpha\frac{F}{D}+(1-\alpha)F
\end{equation}
where $\alpha=(\lambda_B-\lambda_0)/\Delta\lambda + 1/2 = Rz + k$,
here $\lambda_B=4000(1+z)$ is the observed wavelength of the break,
$\lambda_0$ is the central wavelength of the filter, the ``local
resolution'' is $R=4000/\Delta\lambda$ and $k = (4000-\lambda_0) /
\Delta\lambda + 1/2$.  We then have that
\begin{equation}
f = F [ (D^{-1}-1)(Rz+k)+1]
\end{equation}
and
\begin{equation}
z= \frac{1}{R} \left[ \left(\frac{f}{F}-1\right)\left(\frac{D}{1-D}\right)-k \right] \ .
\end{equation}
The error in the redshift roughly depends on the flux measurement
error $\sigma_f$ as
\begin{equation}
\sigma_{z_f} \approx \frac{D}{R(D-1)} \frac{\sigma_f}{F}
\end{equation}
where we have considered that the error in the determination of $F$ is
much smaller than $\sigma_f$.

 Apart from the photometric error, another source of uncertainty in
the redshift estimation is the intrinsic variability of galaxy spectra
around its average, even within such a homogeneous class as LRGs.
\citep{Cool2006, EisensteinLRG}. We can include this in our toy model
as an uncertainty in the $4000$~\AA\ break amplitude $D$. Using the
above formulae we get that
\begin{equation}
\sigma_{z_D} \leq \frac{1}{R(D-1)^2} \sigma_{D} \ . 
\label{eq:template}
\end{equation}
The total uncertainty predicted by the toy model is thus,
\begin{equation}
\sigma_z \sim \frac{1}{R(D-1)^2}\sqrt{ \sigma_D^2+2 (\sigma_f/F)^2} \ .
\label{eq:sigmaz}
\end{equation}

 We can estimate the intrinsic scatter in $D$ to be $\sigma_D=0.1$,
as shown below.  To check the validity of this formula we can use the
photometric observations of LRGs with measured spectroscopic
redshifts.  We have downloaded a LRG catalog with spectroscopic
redshifts from the SDSS website and estimated their photometric
redshifts using the LRG template described below, measuring an average
error of $\sigma_z \approx 0.02(1+z)$. Eq.~(\ref{eq:sigmaz}) clearly
overestimates the error, since at e.g. $z=0.3$, where galaxies have
typically $\sigma_f/F=0.05$, it would predict (using $D=1.8$) an error
from the template variability of $\sigma_{z_D}=0.044$, a photometric
error of $\sigma_{z_f}=0.03$ and a total error of $\sigma_z=0.05$,
about twice as large as the real result $\sigma_z=0.026$. This is not
surprising, since real galaxies have many features which contain
redshift information apart from the 4000~\AA\ break.  Therefore,
although Eq.~(\ref{eq:sigmaz}) can be useful to qualitatively understand the
effects of intrinsic scatter and photometric noise on the photometric
redshift accuracy, it clearly underestimates the precisions which can
be achieved in practice.

  The application of Eq.~\ref{eq:sigmaz} to our set up, with $\sigma_f/F=0.1$ gives $\sigma_z=0.006$, which again is twice what we expect.
In what follows we will perform a detailed simulation to show that it
is feasible to reach the photometric redshift accuracy required for
our experiment ($0.003(1+z)$) under realistic observing conditions, taking into
account the shape of real galaxies, the behaviour of the sky
background as a function of wavelength and lunar phase, and the
expected throughput and efficiency of astronomical instruments.

\subsection{Observational setup and S/N considerations}

\begin{figure}
\epsscale{1}
\plotone{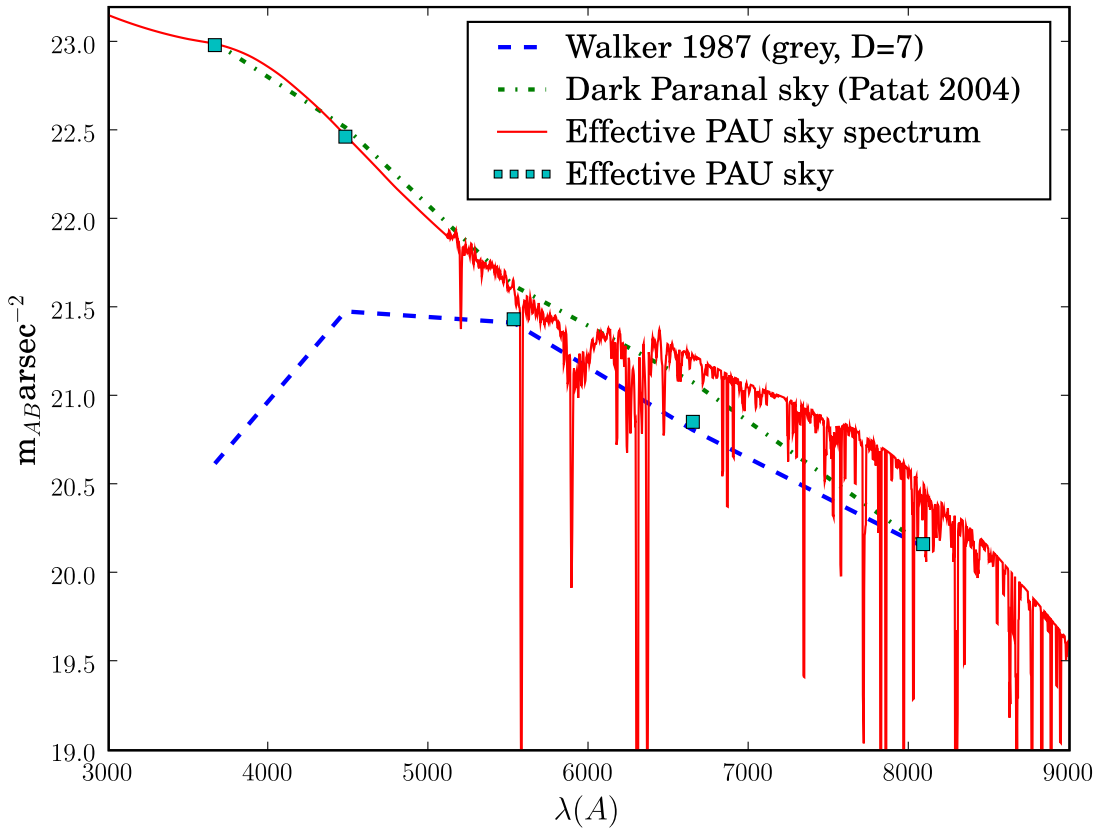}
\figcaption{The sky background assumed for our simulations. We have assumed that
we are able to adapt the choice of filters in our observations to
the moon cycle, observing in the u-band the darkest night, and then
moving towards redder filters as the sky brightness grows. The red,
continuous line corresponds to our expected {}``effective'' sky
spectrum, with the squares showing the equivalent broad band $AB$
magnitudes in the $UBVRI$ filters. The spectrum is normalized to
have the same broad band brightness as the \citet{Patat} measurements
of the dark sky at Paranal for the $U$ and $B$ bands, and the same
as the middle of the cycle {}``gray'' nights from \citet{walker} in
the rest of the filters. 
\label{fig:sky}}
\end{figure}

   In order to simulate the characteristics of the astronomical site
where the PAU observations will be carried out, we assume that the sky
brightness for the dark phase of the lunar cycle is similar to that of
Paranal, as measured by \citet{Patat}. For the middle of the Moon
cycle, or ``gray'' time we use the values of \citet{walker} for Cerro
Tololo. Figure~\ref{fig:sky} shows the assumed sky brightness in the
standard $UBVRI$ broad bands. However, due to the narrowness of our
filters, it is necessary to have a good representation of the small
scale structure of the sky spectrum, for this we use the model optical
spectrum of~\citet{Puxley}, the same used for the Gemini exposure time
calculator.

  We have written an exposure time calculator for this task. To
calculate the full throughput of the system we use the La Palma
atmosphere at 1.2 airmass, two aluminum reflections and the LBNL CCDs~\citep{Holland03}
quantum efficiency curves. We also approximate the throughput of the
filter system using the values of the BARR filters produced for the
ALHAMBRA Survey. The final result is shown in
Fig.~\ref{fig:filters}. We match our results to those of the ING
exposure time calculator, SIGNAL, using the same observational setup.
To reproduce their results, which have been calibrated empirically, we
have to degrade our theoretical estimates by $25\%-10\%$ (which are
basically the values of the empirical corrections they use). We have
checked our predictions with preliminary results from the ALHAMBRA
Survey observations and they agree within $10\%$. 
We have compared the predictions of our simulator with those of 
DIET~\footnote{
http://www.cfht.hawaii.edu/Instruments/Imaging/Megacam\-/dietmegacam.html}, 
the Direct Imaging Exposure Time calculator. DIET estimates $5 \sigma$ point source limiting magnitudes of $g=25.74$ (dark time) and $r=24.94, i=24.49$ (grey time) for 442~s exposures, within $\approx 1$ arcsec$^2$ apertures and 0.8 arcsec seeing. We can scale these results taking into account the effective width of our narrow band filters and the relative collecting mirror areas ($10$ m$^2$ for CFHT vs. $\pi$ m $^2$ for our fiducial telescope), corresponding to limiting magnitudes of $m_{F4982}=23.63$, $m_{F6283}=23.12$ and $m_{F7771}=22.78$. Our predictions shown in Fig 7 are $m_{F4982}=23.77$, $m_{F6283}=22.98$ and $m_{F7771}=23.15$. Most of the differences can be explained by the introduction by DIET of a coefficient which attempts to account for the incorrect measurement of the sky background for very small apertures and which worsens the S/N by a factor 1.22 for faint objects. 

  The simulations below have been carried out assuming that the survey
will use a dedicated 2m-class telescope, with an effective area of
$\pi\textrm {m}^{2}$, and a camera with a 6~deg$^2$ FoV. The results
will remain qualitatively valid as long as the {\it etendue} of the
final observational setup is roughly the same.
Most likely the observations will be carried out in drift-scan mode. Since there
is no need to change instruments, we expect that the observing
efficiency will be very high and that only a maximum of two CCD
readouts per filter will be carried out.
Assuming that the useful time will be similar to that in 
Calar Alto~\citep{sanchez2007} and that the moonlight will prevent us from taking data during 3 nights per Moon cycle, the number of useful hours per year amounts to 1930. Leaving some room for unforeseen incidences, we assume that the total number of hours of exposure time per year will amount to 1800. For a survey area of 8,000 sq.deg, and with a 6 sq. deg. camera in a period of 4 years we expect to be able to expose each field a total of 5.4hrs, or 19440s.

The best way of measuring accurate colors for photo-z is using
relatively small isophotal apertures~\citep{Benitez2004} which
maximize the S/N of the color measurements, despite the fact that such
an aperture leaves out a large amount of the flux, and they are
therefore not optimal for other scientific purposes. In our S/N
estimations we assume that we will use $2$ arcsec$^{2}$ apertures,
which enclose about $40\%$ and $64\%$ of the flux of respectively a
$z=0.2$ and a $z=0.9$ $L_\star$ galaxy.  In the next subsection we
explain how we calculate these corrections.

\begin{figure}
\epsscale{1}
\plotone{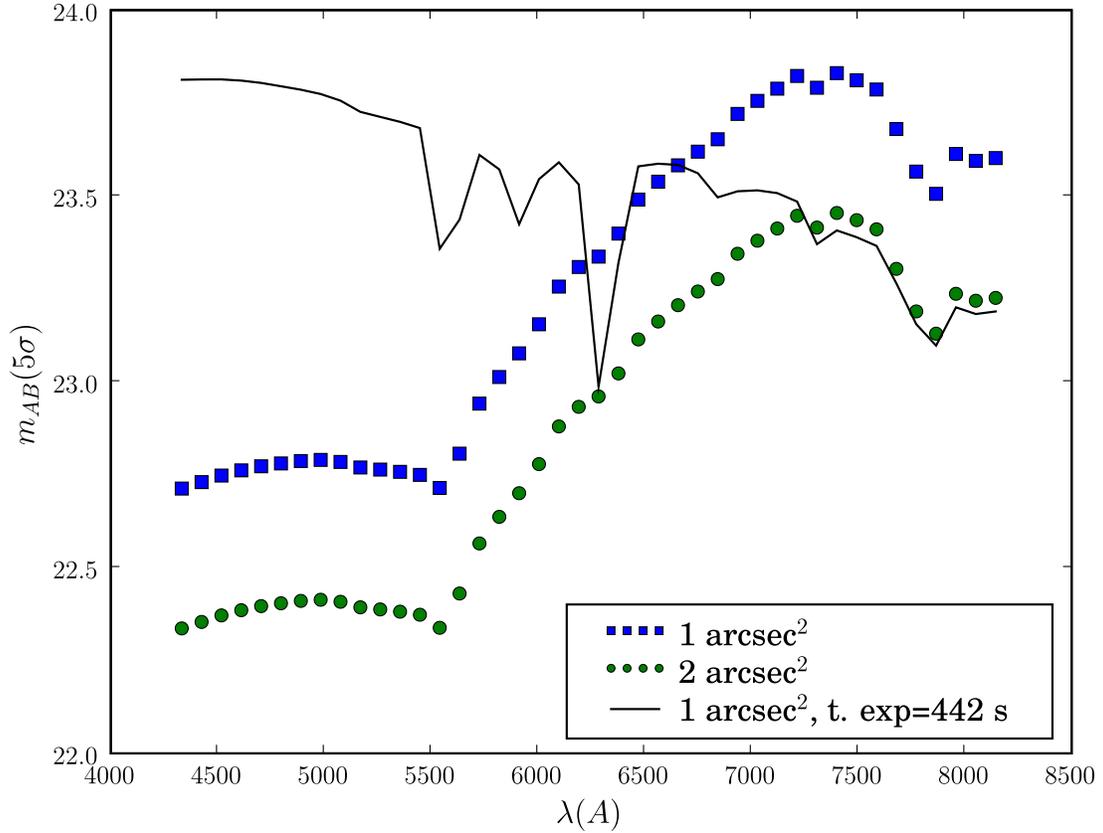}
\figcaption{Expected limiting magnitudes for PAU-BAO. The squares represent the 
$5\sigma$ magnitude limits within a 1 sq. arcsec aperture, the circles within a 2 sq. arcsec aperture and 
the continuous line is the $5\sigma$ magnitude limit which would be reached 
if we divided the total exposure time of 19440s equally among all the filters.
\label{fig:mag5}}
\end{figure}

\begin{figure}
\epsscale{1}
\plotone{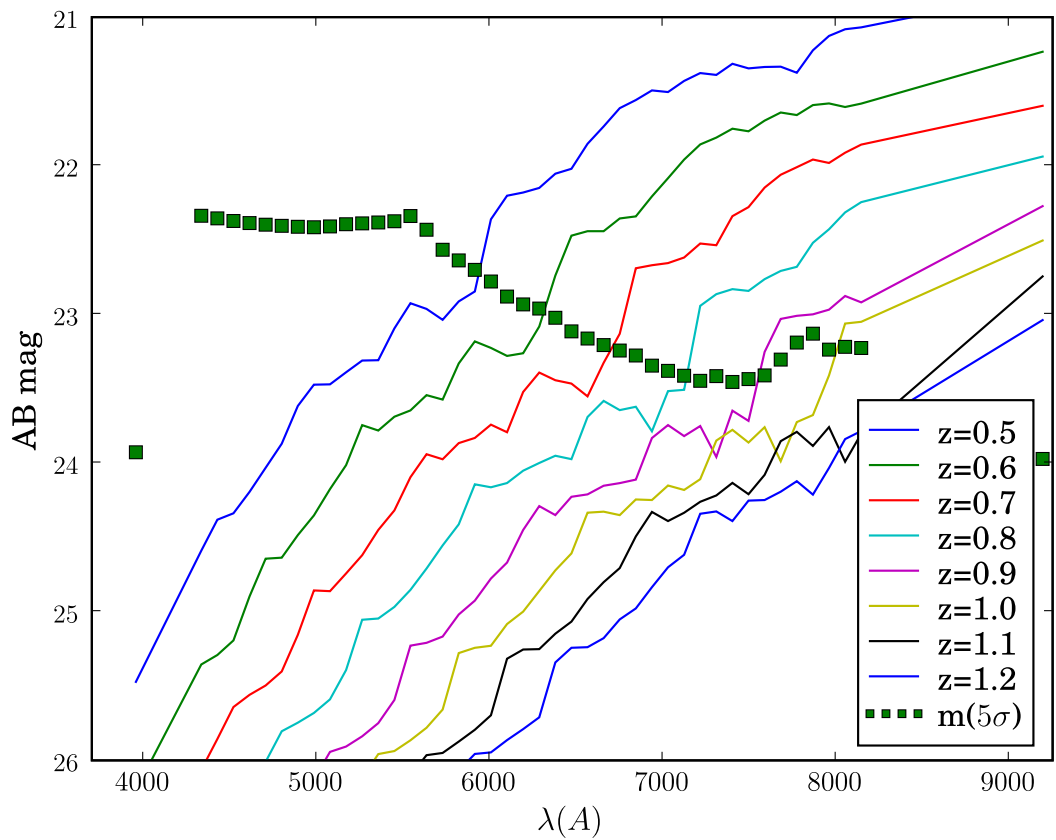}
\figcaption{The expected 5-$\sigma$ limiting magnitudes for point
sources (squares) and the observed spectra of a $L_{\star}$ red galaxy
at different redshifts (without taking into account spectral evolution, but taking 
into account aperture corrections). Note
that we are able to catch the rest frame 4000~\AA~ break with
enough filters on both sides up to z=0.9.
\label{fig:mag}}
\end{figure}

  A crucial question is how to divide the exposure time between the
different filters.  At each redshift, we identify the filter which
corresponds to the $4150\AA$ rest frame region, and try to detect a
LRG $L_\star$ within a $2$ arcsec$^{2}$ aperture at that redshift with
at least a S/N of 10. We set a minimum exposure time of $120$s, and
adjust the maximum exposure time in each filter so that the total is
below 5.4 hours. The resulting exposure times are $< 120$s for filters
bluer than $F5446$, and increase until they reach the maximum exposure
time of $861$s for $F7307$ and redder filters. 
The resulting $5\sigma$
limiting magnitudes are plotted in Fig.~\ref{fig:mag5} and
Fig. ~\ref{fig:mag}.

\subsection{Intrinsic galaxy variability}

  As Eq.~(\ref{eq:sigmaz}) shows, it is necessary to understand the intrinsic
variability of LRGs in order to estimate the photometric redshift
accuracy achievable with them. It is well known (Eisenstein et
al. 2003, Cool et al. 2006 and references therein) that LRG galaxies
(with $L>2.2L_{\star}$) are a remarkably homogeneous class. At a fixed
redshift, they form a red sequence, which varies slowly and regularly
with absolute magnitude and environment. LRG galaxies in the red
sequence present a scatter of only a few percent in the color defined
by a pair of filters spanning the $4000\AA$ break. Therefore, if we
know the absolute magnitude of a LRG and the richness of its
environment, we can predict its broad band colors with a precision of
at least $\sigma_{g-r}\approx0.035$ (Cool et al. 2006).

It is not clear however, which are the actual variations in the
spectral shape of LRGs behind this broad band scatter. The SDSS
spectrophotometry is not good enough to accurately characterize this
phenomenon, since its precision (about 0.05 mags in g-r colors, according to the SDSS
web site and Adelman-McCarthy et al.~(2008)) is of the same order or even
larger than the intrinsic color variation of real galaxies.  In
addition, the errors in the spectrophotometry are bound to be highly
correlated and will be much worse at certain wavelength regions, like
sky lines. The section of the SDSS website which describes the quality
of the spectrophotometric calibration shows that below $4000\AA$ the
flux calibration error can be as large as $10\%$ .

  Since it is not feasible to use the SDSS spectral information, we
have therefore decided to use a different approach to characterize
this intrinsic spectral variability.  Eisenstein et al.~(2003) split the
spectra of LRG into different classes and samples, and looked at the
differences amongst them using Principal Component Analysis (PCA).
They showed that most of the variation between these average classes
can be ascribed to a single spectral component. It is therefore
reasonable to assume that the intrinsic variation for galaxies of each
class, responsible for the red sequence scatter described above, can
be modeled approximately using the same PCA component.  We have
therefore generated a mock galaxy sample with $L>L_{\star}$ at $z=0.16$
using the red sequence described by Cool et al.~(2006), and the
luminosity function described in Brown et al.~(2007), and fit their
$g-r$ and $r-i$ colors using the average template of Eisenstein et
al.~(2003) and the first PCA component (shown in Fig.~8 of Eisenstein's
paper).

   The reason to limit ourselves to those two filters is that
Eisenstein et al.~(2003) only provide spectra in the $3650\AA-7000\AA$
wavelength range, which does not include other filters. The comparison
with the average Cool et al. 2006 colors show that we have to slightly
correct Eisenstein's average template to adapt it to the observations,
subtracting the first PCA component multiplied by 1.74, and that the
required variation of the amplitude of the PCA component needed to
explain the intrinsic scatter around the red sequence is approximately
1.8 times that necessary to explain the variation of LRGs with
redshift, magnitude and environment.  The average D4000 is 1.81, and
the rms around this average value is 0.104.  Therefore the $4000\AA$ 
break amplitude seems to display an intrinsic scatter of $6\%$ in real
galaxies.

\subsection{Input early type catalog}

  To describe the early type galaxy population we use the luminosity
functions described in Brown et al.~(2007). We populate a 10~sq.~deg.~area
with galaxies following this distribution and exclude those which
are fainter than $L_{\star}$ (although it is obvious that many of them
will be detected as well) and fainter than $I_{AB}=23$.

\begin{figure}
\epsscale{1}
\plotone{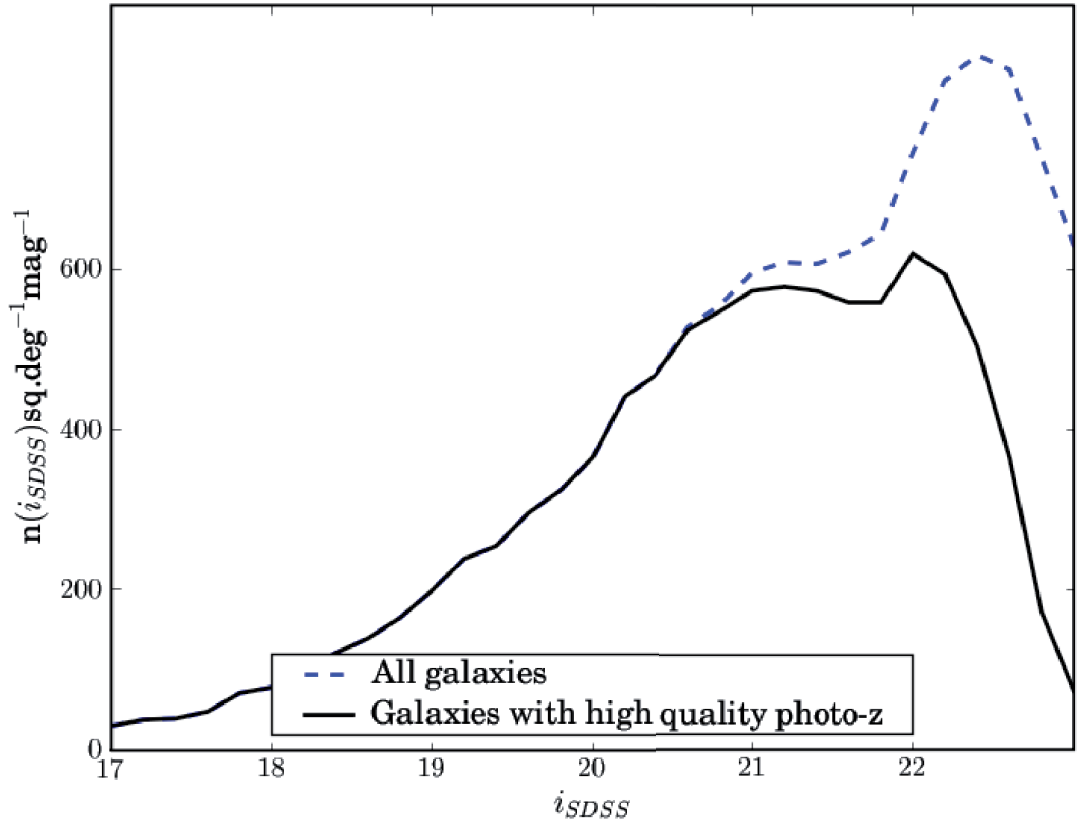}
\figcaption{Differential number counts distribution of $L>L_{\star}$ red galaxies.
\label{fig:nc}}
\end{figure}

\begin{figure}
\epsscale{1}
\plotone{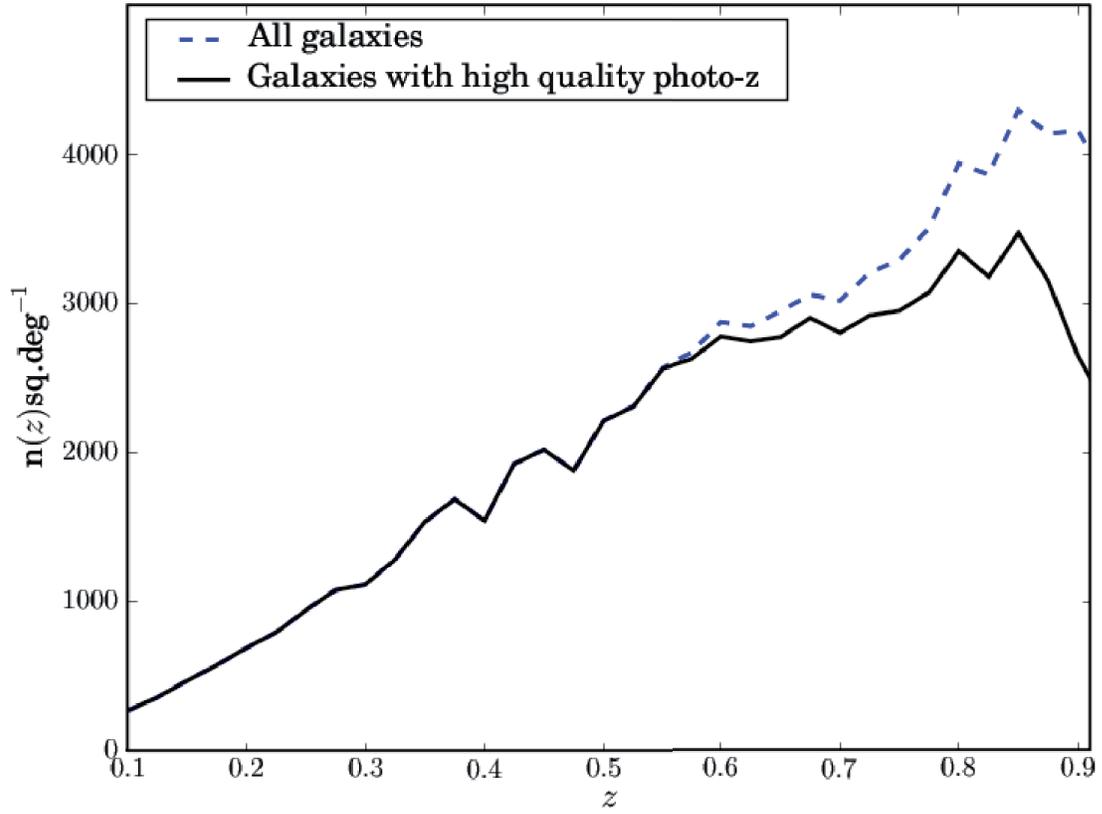}
\figcaption{Redshift distribution of $L>L_{\star}$ red galaxies.
\label{fig:nz}}
\end{figure}

 LRGs are extended objects, and we have to calculate which is the
fraction of the total flux which falls within our reference 2~sq.~arcsec aperture. 
For this we use the data on galaxy sizes and
their evolution provided by Brown et al.~(2007), assuming, as they do,
that galaxies can be well represented by a de Vaucouleur profile. The
correction ranges from $1.5$ mags at z=0.1 to $\approx 0.5$ for
$z>0.7$.  The resulting differential numbers counts and redshift
distribution are plotted with dashed lines in Figs.~\ref{fig:nc} and \ref{fig:nz} (note
that we plot the magnitudes corresponding to a 2~sq.~arcsec aperture).

\subsection{Results}
  To simulate our observations we will redshift and integrate under
the corresponding filter transmissions a spectrum resembling a typical
LRG galaxy, and then try to recover its redshift using a Bayesian
photometric redshift method (BPZ, described in~\citet{Benitez2000}).
Obviously in the real world we will not use a single template for all
LRGs between $0 < z < 0.9$: their spectra are known to vary with
redshift and luminosity (Eisenstein et al. 2003; Cool et al. 2006).
However, as that paper shows, the variation is smooth and easy to
parameterize.  This is confirmed by HST very deep observations of
galaxy clusters, where the scatter around the red sequence remains
small ($\sim 0.03$) up to $z=1$ and higher \citep{Blakeslee2003}.

 We assume that we will be able to split our LRGs into subsamples such
that for each of them we can define an empirically calibrated template
(using a technique similar to that of \citet{Budavari2000} or
\citet{Benitez2004}) which correctly represents the average galaxy
colors for that galaxy subsample.  Using standard photo-z
techniques~\citep{Benitez2000} we reasonably expect to be able to
determine the redshift and spectral type of our galaxies in a
preliminary pass to within $0.01(1+z)$.  This is already being done
for the ALHAMBRA survey~\citep{Benitez2008,Moles2008}.  Thus, for each
galaxy we will have a preliminary estimate of its redshift to within
$\sigma_z \sim 0.01(1+z)$ and its absolute magnitude to with
$\sigma_M\sim 0.15$ (within the redshift interval $z<0.9$).  This
ensures that we can pin down the required template for each galaxy
with large certainty (the error in the absolute magnitude corresponds
to an intrinsic color variation of only 0.004~mag, much smaller than
the expected scatter around the sequence at each redshift,
0.03-0.04~mag).

  The LRG template corresponds to $z<0.5$ galaxies, and one may wonder
if the results obtained with this template are representative of
higher redshift LRGs. Homeier et al.~(2006) have measured the
$V_{606}-I_{814}$ 
colors of a pair of clusters at $z=0.9$ with the
Advanced Camera for Surveys aboard HST. They measure a red sequence
that is only $0.09$ bluer than the colors predicted by our LRG
template which illustrates the small amount of color evolution
expected to $z<1$ and shows that the results obtained with our LRG
template should be similar to those obtained with real LRG templates
at higher redshifts.

 In our simulation we generate the galaxy colors using a combination
of the average Eisenstein et al. 2003 template, corrected as mentioned
above and the first PCA component, multiplied by a coefficient with a
Gaussian distribution of rms=1.8. This first PCA component scatter
represents well the real broadband scatter of galaxies observed by
Cool et al. (2006) for $L>2.2L_{\star}$ LRGs. We extrapolate this
scatter to $L>L_{\star}$. We also add a $2\%$ noise to represent the
expected scatter in the zero point determination across the survey

    Then we calculate photometric redshifts using the Bayesian
photometric redshift method implemented in the BPZ code and a single
LRG reference template. We have also tried a template library with 11
templates, formed by linear combinations of the LRG template and the
first PCA component encompassing $\pm 3\sigma$ variations and the
results are basically the same. For simplicity we quote the results
obtained with only one LRG template.

\begin{figure}
\epsscale{1}
\plotone{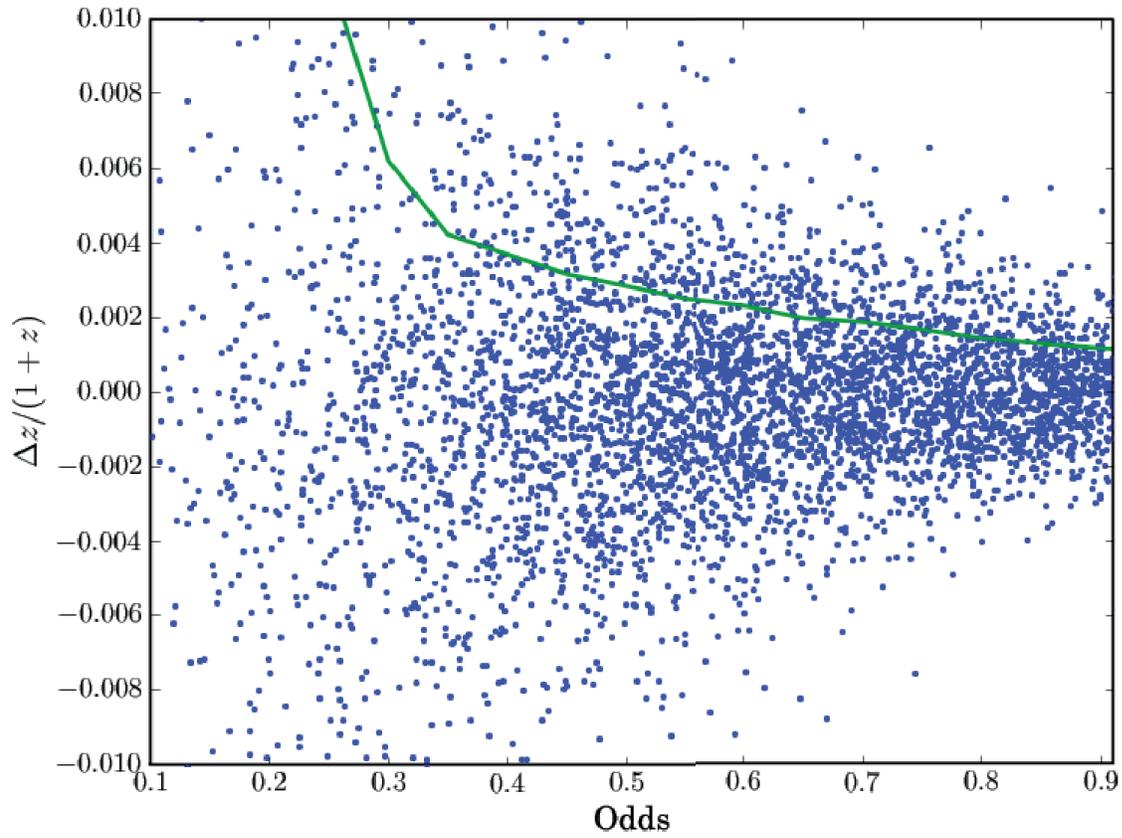}
\figcaption{Photometric redshift error as a function of the Bayesian odds. Note that a cut at odds $= 0.55$ 
eliminates most of the objects with high redshift errors. For the sake of clarity,
only one in every five points is plotted. The solid line corresponds to the rms of $\Delta z/(1+z)$
for each value of the odds.
\label{fig:odds}}
\end{figure}

   With a single template, there is no point in using a prior, but the
Bayesian framework still remains useful: it produces the so called
``odds'' parameter, a highly reliable quality indicator for the
redshift estimate. In Fig.~\ref{fig:odds} we plot the scatter diagram
corresponding to the quantity $(z_{phot}-z_{s})/(1+z_{s})$, where $z_s$ is the true
redshift, as a
function of the odds parameter, together with the rms corresponding to
each value of the odds. We can see that if we exclude the objects with
low values of the odds parameter we get rid of most of the redshift
outliers.  The effectiveness of this technique has been often
validated with real data \citep{Benitez2000,Benitez2004,Coe2006}. Note
that using a cut in $\chi^2$ does not work well to eliminate outliers,
as it was shown by~\citet{Benitez2000}.

\begin{figure}
\epsscale{1}
\plotone{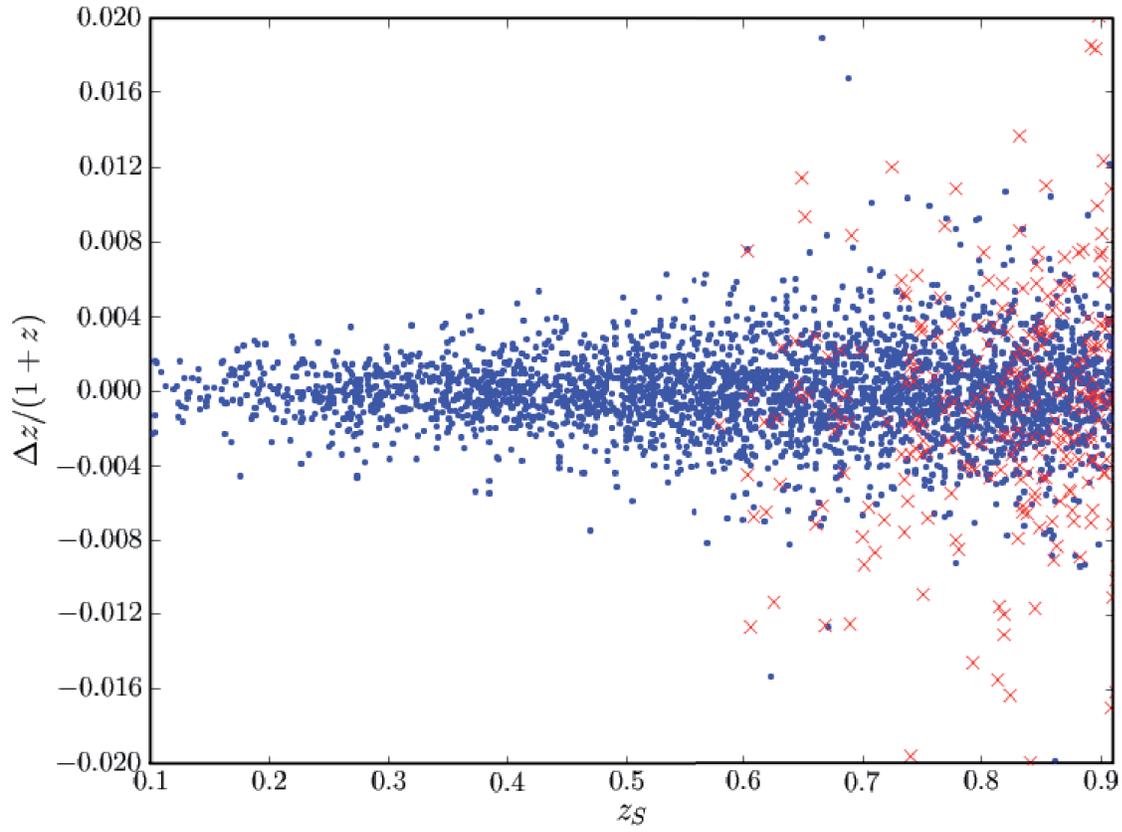}
\figcaption{Scatter plot comparing the normalized difference 
between the photometric redshifts and the ``true'' input redshifts $z_S$. The red points are eliminated
by the odds~$<0.55$ cut. For the sake of clarity, only one in every five points is plotted.
\label{fig:scatter}}
\end{figure}

 We thus proceed to eliminate the objects with odds $< 0.55$ from our
catalog.  Figure~\ref{fig:scatter} shows the scatter diagram for
$(z_{phot}-z_{s})/(1+z_{s})$ now as a function of the real redshift,
$z_s$.  Once the odds cut is applied there are no large outliers.

\begin{figure}
\epsscale{1}
\plotone{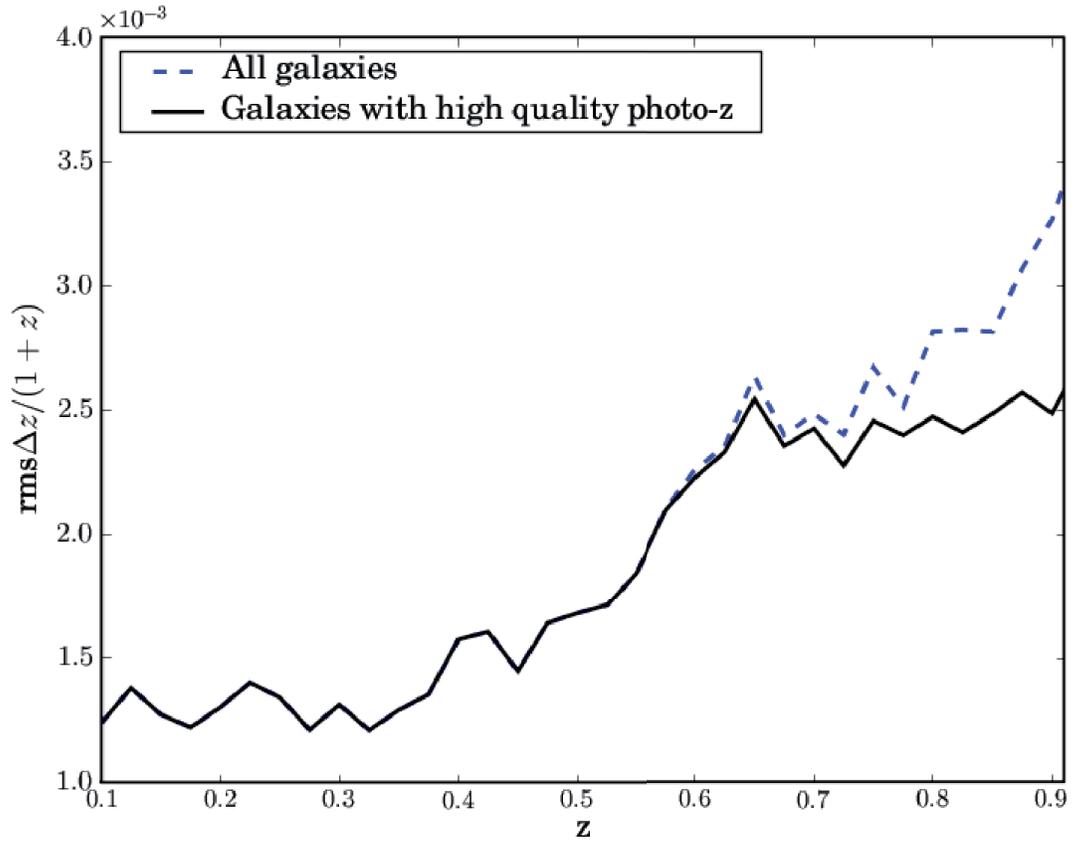}
\figcaption{Photometric redshift error as a function of redshift, for all $L>L_{\star}$,$I<23$ red galaxies, 
and for the subset with high quality photo-z.
\label{fig:res}}
\end{figure}

 The resulting redshift and number counts distributions are plotted as solid lines in
Figs.~\ref{fig:nc} and \ref{fig:nz}. In Fig.~\ref{fig:res} we plot the
resulting accuracy as a function of redshift.  We are safely below the
$0.003(1+z)$ limit for all our redshift range.  Finally, in
Fig.~\ref{fig:nv} we plot the number density of all the galaxies, and
of those with high-quality photo-z as function of redshift.
These figures show that we have a spatial density of $\bar{n} >
10^{-3}$~(h/Mpc)$^3$ in the redshift range $z<0.9$. Since $P(k) >
10^{4}$~(Mpc/h)$^3$ for LRGs (see eg.~Fig.~4 in~\citet{Tegmark06})
and $k<0.2$~h/Mpc, we will have $\bar{n}P(k)>10$ for the $k$ range of
interest for BAO, so that, according to Eq.~(\ref{eq:deltaP}), shot
noise will be negligible.

  Finally, there are two caveats to consider. First, there are no
spectroscopic data with good enough spectrophotometric calibration for
LRGs in the redshift range of interest. We can therefore only estimate
the intrinsic variation of the galaxies from the data available.  We
have assumed that it will behave similarly to the variation among LRG
types described by Eisenstein et al.~(2003). Second, the PCA study only
covers the $3650\AA-7000\AA$ range, and we assume that there is no
template variation outside this range. We feel that this is justified
since most of the redshift information for the galaxies is in practice
contained in this interval, especially at high redshift.

\subsection{Comparison with a spectroscopic survey}
A typical multi-fiber spectroscopic survey with about 1000 fibers and
a resolution $R\sim2000$ in a telescope similar to the one we are
assuming here (2-meter class, about 6-deg$^{2}$ FoV, etc.) will reach
up to a magnitude $i<20$ in about 2-hour-long exposures~\citep{BOSS},
assuming the transmission of a good optical spectrograph and low
readout noise.  This allows covering in a year close to
4000~deg$^{2}$ with $0.1<z<0.8$ for LRGs, or about 2.5~(Gpc/h)$^{3}$
per year. In our PAU approach, with our 300-900~s (depending on
the band) exposures, we can cover about 2000~deg$^{2}$ per year with
$0.1<z<0.9$ for LRGs, which translates to about 2~(Gpc/h)$^{3}$ per
year, however with higher galaxy density. This results in $\bar{n}P(k)
>10$ at the relevant scales (see Eq.~(\ref{eq:deltaP})), while for a
spectroscopic survey similar to~\citet{BOSS}, with about 1000
fibers in a 6~deg$^{2}$ FoV, one can only reach
$\bar{n}P(k)\sim1$. Putting volume per year and galaxy density
together, for an equal-time survey one gets
\begin{equation}
\frac{\left(\Delta P/P\right)_{\mathrm{PAU-BAO}}}
{\left(\Delta P/P\right)_{\mathrm{spect}}}=\sqrt{\frac{2.5}{2}}\ 
\frac{1+1/10}{1+1/1}\sim0.6\label{eq:compa-PAU-BOSS}
\end{equation}
For the radial modes, one further needs to take into account the
slight degradation in information that affects the PAU measurement
with its $\sigma(z)=0.003(1+z)$. 

Furthermore, in the imaging survey one gets many more galaxies than the
LRGs. A preliminary study for the whole galaxy population obtains a
good photometric redshift determination, $\sigma(z) \approx 0.01
(1+z)$, for a large number of them (over 200 million). These galaxies
would deliver a constraint on the BAO scale of similar power than the
one from LRGs (although correlated, since both galaxy distributions trace 
the same underlying density fluctuations), so that the combination of both 
would improve the sensitivity, and could serve as a cross-check on systematic errors.
\subsection{Calibration Requirements}
We present here some general considerations to give an idea of what
level of photometric and spectroscopic calibration is required to
measure the BAO scale with PAU.  In the next section we will address
the issue of whether these requirements can be met in practice.  We
split this section into photometric and photo-z requirements.
\subsubsection{Photometric Calibration}
The magnitude of a galaxy that we measure in the survey, $m_{O}$, is
the sum of the true magnitude $m$, plus a random statistical error
that arises from photon and detector noise, $e_{mr}$, plus a
systematic error $e_{ms}$. The systematic error arises from a variety
of effects. For example, variations across the survey of the exposure
time, mean atmospheric absorption and sky background; non-uniformity
of galactic dust absorption and inaccuracies in its correction;
variations in the instrument/detector efficiencies through the
duration of the survey. All these effects are assumed to have been
corrected for through calibrations with standard stars and flat
fielding corrections. But inevitably, every correction has an error
which contributes to $e_{ms}$. While the random statistical errors of
any two galaxies are uncorrelated, the systematic errors in the
magnitude over the survey have a correlation function
$\xi_{ms}(z,\theta,g)$ that is likely to depend on redshift and
angular separation $\theta$, as well as galaxy properties $g$ (e.g.,
luminosity, morphology or color).

The random statistical errors have an effect that is reduced as the
number of galaxies is increased. Generally the galaxy number shot
noise will be larger than the error introduced by random errors in the
apparent magnitude, as long as these magnitude errors are not very
large. So if the number of galaxies that is observed is large enough
the statistical errors should be small compared to the uncertainty in
the correlation function due to cosmic variance. The systematic
errors, however, do not go down with the number of galaxies observed.

For a flux limited survey, a magnitude calibration covariance across the
sky $\Delta_{m}(\theta)$ will result in angular density fluctuations
$\delta(\theta)$. If we take the number of galaxies brighter than
magnitude $m$ to be $N(<m)\simeq10^{\alpha\; m}$ (typically
$\alpha\ln{10}\simeq1$), then a magnitude error translates into a
number density fluctuation error:
\begin{equation}
\delta\simeq\alpha\;\ln{10}\;\Delta_{m}\ .
\label{deltam}
\end{equation}
We can decompose the calibration error field in the sky into spherical
harmonics. We would like the resulting spectrum of calibration
errors $C_{l}^{m}$
\begin{equation}
C_{l}^{m}=2\pi\int_{-1}^{1}\mathrm{d}\cos\theta\,\Delta_{m}(\theta)\, P_{l}(\cos\theta)
\end{equation}
to produce errors in the angular power spectrum $C_{l}$ which are
smaller than the sampling variance errors in $C_{l}$:
\begin{equation}
C_{l}^{m}<\frac{\Delta C_{l}}{{\alpha\;\ln{10}}}\simeq\frac{C_{l}}{{\alpha\;\ln{10}\sqrt{f_{sky}(l+1\
/2)}}}
\end{equation}
We will assume that angular clustering will be sampling variance
rather than shot-noise variance dominated. We also assume Gaussian
statistics. The corresponding errors in the correlation function $\Delta_{m}(\theta)$
are: \begin{equation}
\Delta_{m}(\theta)=\sum_{l}\frac{2 l+1}{4\pi}C_{l}^{m}~P_{l}(cos\theta)
\label{eq:req}\end{equation}

The BAO scale projects at angles between $3.7$ and $1.7$ degrees
for redshifts between $z=0.4$ and $z=1.0$ (smaller redshifts cover
a negligible volume). Unfortunately the field of view (FoV) of the
planed PAU camera plans to cover very similar angular scales. We therefore
need to be careful about calibration on the FoV. At angular
scales $3.7$ and $1.7$ degrees, the requirement in Eq.(\ref{eq:req})
translates into rms correlated calibration errors smaller than $2\%$
to $3\%$ (ie $0.02$ and $0.03$ rms magnitude errors) in units of $(b/2)/(\alpha~\ln{10})$ for $f_{sky}=0.2$. 
This is for the whole (flux limited) sample (mean $z\simeq0.7$). These constraints become
looser when we split the sample into redshift bins because the amplitude
of clustering increases as we reduce the projected volume. The detailed
constraints are shown in Fig.~\ref{fig:d_m}.
\subsubsection{Selection effects on $\xi_2(r)$}
Another angle to look at possible photometric calibration effects is
to assume that different systematics on the galaxy density
fluctuations will act as multiplicative correction over the galaxy
density at a given position in the sample. We will assume that this
type of error is uncorrelated to the galaxy clustering so that:
\begin{equation}
\xi_{obs}(r) = \xi(r) + \xi_e(r)
\end{equation}
where $\xi(r)$ is the true correlation and $\xi_e(r)$ is the
correlation due to systematics in selection and calibration. To see
how this could affect the BAO scale measurement we model the true
correlation around the BAO scale as a Gaussian peak of width
$\sigma_0$. We further assume a generic power-law $\xi_e(r) \propto
r^{-\beta}$ for the error around the BAO scale.  The relative shift in
the BAO scale can be found by Taylor expansion around the peak:
\begin{equation}
\Delta_{BAO}=  \beta ~ { \xi_e(r_{BAO}) ~ \sigma_0^2
\over{\xi (r_{BAO}) ~ r_{BAO}^2}}.
\end{equation} 
This requires an amplitude of $\xi_e(r_{BAO})< 0.002$ if we want a
shift in the peak $\Delta_{BAO} < 1\%$ and $\beta \simeq 2$. We have
used here $\sigma_0 \simeq 15$ Mpc/h, $r_{BAO} \simeq 100$ Mpc/h, and
$ \xi(r_{BAO}) \simeq 0.01$ from Fig.\ref{fig:xi2}. 
This corresponds to a $20\%$ error in the correlation at the BAO scale, 
and about $\sqrt(\xi_e)=4.5\%$ error on density fluctuations.

We have also tested the above
calculations directly in simulations, by adding $\xi_e(r)$ and
recovering the BAO scale.

Using Eq.~(\ref{deltam}), this value of $\xi_e(r_{BAO})< 0.002$
corresponds to $\Delta m < 0.05$ (in units of $\alpha\;\ln{10}$ with
$N(<m)\simeq10^{\alpha\; m}$). Thus, with very different assumptions
we reach the same conclusion on the requirement on photometric
accuracy of around $5\%$ on the BAO scale.
\subsubsection{Photo-z bias}
Systematic errors in the radial direction (photo-z biases) also need
to be under careful control. At any given redshift, we would like the
mean in the photo-z measurements to differ from the true redshift by
less than $1\%$ (the target in $\Delta_{BAO}$ accuracy) in the radial
BAO distance, ie $\sigma_r \simeq 1 $Mpc/h, which corresponds to:
\begin{equation}
\Delta_z = \sigma_r~H(z)/c  \simeq 5 \times 10^{-4} ~  (z=0.8)   
\end{equation}
where the numerical value corresponds to $z=0.8$ and $\Omega_m=0.2$.
This is about an order of magnitude better than the statistical error
at the same redshift, i.e.~$\sigma_z \simeq 0.003 (1+z) \simeq 5 \times
10^{-3}$.

Note that this is a conservative approach because we need the
$\Delta_{BAO}$ accuracy as measured by galaxy density fluctuations and
not by the absolute distances to the galaxies.  The former will probably
result into a weaker constraint for $\Delta_z$.
\subsection{Calibration Plan}
As a summary for the above requirements, we need relative calibration
to be better than about $3\%-5\%$ to avoid systematic effects on
density fluctuations to dominate over intrinsic fluctuations on the
BAO scale. On top of this we would like to have a bias in the photo-z
scale to be below $1\%$ on radial measurements of the BAO scale.

In terms of global photometry, it has now been shown that a
homogeneous global relative calibration below $2-3\%$ accuracy is
possible in current and future surveys (e.g., \cite{Sterken07} and
references therein). The large field of view required by the PAU survey
and the drift scanning strategy will both help in the provision of
standard calibration techniques, such as done in SDSS.  We will also
need to use a set of calibrated standard spectra of stars (or
galaxies) to monitor and correct for relative color bias between
narrow bands.

Apart from these "classical" techniques, it seems to be possible to
use the observed colors of galaxies with spectroscopic redshifts as a
photometric calibrator. In HST´s Ultra Deep Field \citet{Coe2006} have
been able to calibrate the NICMOS zero-points using the comparison
between predicted colors using the templates of \citet{Benitez2004}
and $\sim 50$ spectroscopic redshifts.  Similar techniques have been
used for the COSMOS field \citep{Capak2007} and are being applied to
ALHAMBRA \citep{Moles2008}.  Further work with the later survey will
help refine our calibration redshift requirements.

Independently of the exact photo-z method finally used for the survey,
it will be equivalent to defining a function $z=f(p,C,o)$, where $p$
are a set of parameters describing the function, $o$ are a set of
observables like the approximate redshift of the galaxy (determined
with standard photo-z techniques), its luminosity, size or environment
density, and $C$ are the observed colors. The LRG population under
study is relatively homogeneous and its changes with redshift and
magnitude can be described with a very compact set of parameters
\citep{EisensteinLRG}.  If we determine the parameters $p$ with enough
precision to reach a redshift error $\sigma_z$ over the whole range
under consideration, then, provided that the parametrization is
flexible enough to adapt itself to the observed redshift/color
relationship of galaxies, the systematic zero point error, averaged
over all the galaxies, will be equivalent to $\sigma_z /\sqrt{N_c}$
where $\sigma_z$ is the rms redshift error and $N_c$ is the number of
calibrators. We expect to have a calibration set with several thousand
galaxies, easy to reach with 10-m telescopes as the GTC, and therefore
will have negligible redshift bias.
\section{The Survey Instrument}\label{sec:base} 
The approach of the PAU project is to use known and proven technologies
to build a large field of view (FoV) camera and mount it on a telescope 
that is optimized for the survey.

The simulations presented in previous sections use a baseline concept of 
camera plus telescope with a large 
{\it etendue}, $A\Omega\approx 20$~m$^2$deg$^2$. This {\it etendue} is
achieved by means of a telescope with a 2-m effective aperture and a
camera with 6~deg$^2$ FoV. A total of 42 filters is considered, each
one having a width of 100~\AA ~in wavelength. The full filter system 
covers a range that goes from $\sim$4000~\AA ~to $\sim$8000~\AA, completed
by two broad-band filters similar to the SDSS $u$ and $z$ bands.

In this section, we present the main ideas of a possible implementation of 
such a system. The goal is to show the feasibility of the telescope/camera 
system, and not to present a complete design.

\subsection{Optics}\label{sec:optics} 
Achieving a many-filter, very-large-area survey in a relatively 
short time and the need for a rather large value of the {\it etendue}
demand a very large FoV. The depth is not a major concern since
the targets are bright galaxies in all the surveyed redshift 
range. Therefore, the field of view is the main driver of the optical 
design.
 
To give some quantitative estimates, with a telescope of 2.3 m 
aperture, a pixel size of 0.4\char`\"{} and state of the art
detectors, it is possible to reach S/N $\sim$ 5 for a star of 
m(AB) = 23.5 in $\sim$300 sec in all the spectral range bluer 
than ~7500~\AA. Since the survey should cover, as 
argued before, an area of at least 8000 square degrees, a FoV of 
6 sq.~deg.~is needed to be able to perform the survey in 
4-5 years.

It is important to notice that we do not intend to use detailed 
information on the morphology/shape of the objects, which relaxes 
the requirements on the image scale. This is the reason to choose 
a rather modest plate scale that translates 15 $\mu$m pixels into 
26 arcsec/mm.

These FoV and plate scale are the basic requirements for a telescope 
that will be dedicated to the survey until its completion. These optical 
requirements for such a large FoV telescope and the corresponding 
panoramic CCD camera are demanding but they appear feasible. 

The next optical elements are the filters. They are intended to have
transmission curves with very sharp limits and minimal wavelength
overlap, very similar to the filters in the ALHAMBRA 
survey~\citep{ALHAMBRA, Moles2008, Benitez2008} but 100~\AA-wide. 

The location of the filters in the path to the detectors can affect
the final efficiency of the system. Two options are being considered:
attaching the filters directly over the CCDs or on plate holders that 
could be interchanged. The first option is mechanically simpler but 
reduces the survey flexibility. The second one allows the optimisation
of the exposure times using different sets of filters depending on the
moon-phase or any other external constraint, but its practical 
implementation is more demanding. The final decision will
be taken when all the practical aspects of the survey, such as observing
mode and calibration strategy are fixed.

\subsection{Focal-Plane and Observing Strategy}
\label{sec:focal} 

The baseline concept for the camera is a large mosaic of CCDs
covering the 6 deg$^2$ FoV. The scientific goals can be
reached with pixels of 0.40\char`\"{}. Since most of the current astronomical 
large CCD detectors have pixel scales of 15 microns, we need a camera
of around 500~Mega-pixel or a number of 2K$\times$4K CCDs that ranges
in between 60 and 80, including a few CCDs for focusing and
guiding purposes. 

The baseline CCDs under consideration for the PAU camera are the
fully depleted, high-resistivity, 250 micron thick devices developed
by the Lawrence Berkeley National Laboratory (LBNL)~\citep{Holland03}. These 
CCDs ensure
a very high quantum efficiency in the red zone of the wavelength region
covered by PAU. However, the different possibilities of optimisation 
have some impact in the focal plane instrumentation. There is the possibility 
of having two different types of CCDs covering different regions of the focal 
plane, in direct correlation with the filters, in order to maximise
the sensitivity in the whole wavelength range. Thin blue-sensitive
CCDs correlated with blue filters and thick red-optimized CCDs correlated
with red filters. Several suppliers of CCDs are available for thin 
blue-optimized CCDs. A final decision for the focal plane intrumentation will 
be taken considering the global optimisation of the survey. 

The camera vessel will need to contain a liquid nitrogen reservoir
to maintain the focal plane cold for several hours, ensuring stability
and not compromising the efficiency of the observations. We will also 
investigate the choice of cryo-coolers or pulse tube coolers, taking into
account the mechanical requirements that impose a stable precision of a 
few microns on the positioning of the CCDs in the focal plane.

The best observing strategy for a project of these characteristics is 
the time-delay-and-integrate (TDI) drift scanning mode. Drift scanning
is a powerful imaging technique in which the telescope is kept stationary
and one lets the sky image drift across the CCDs. Normally, TDI mode is 
operated at sidereal rate, and the lines of the CCD must be
read in perfect synchronisation with the movement of the sky at the focal 
plane. In this way, long continuous strips of the sky are imaged and large
fields can be explored automatically, which makes this observing strategy 
particularly well-suited for large surveys. Therefore, it is considered as
the baseline strategy for PAU.

\subsection{Front-End Electronics}
\label{sec:electro} 

The readout of a focal plane of the size of the PAU camera is a 
challenge. One of the most attractive options to read out such a large 
number of CCDs is to use the open system MONSOON~\citep{monsoon}, 
developed by the 
instrumentation division of NOAO (National Optics Astronomical 
Observatory, supported by the US NSF). A custom-made system or a commercial 
one, such as the Leach controller~\citep{leach}, are also being considered. 

MONSOON is a generic readout system which consists in three kind of 
boards: the Acquisition Board, responsible for the bias voltage generation 
and the digitization of signals coming from the CCDs, the Clock 
Board, responsible of the clock signal generation needed to read out the 
CCD, and the Master Control Board, responsible of the event building and the 
data transmission to the DAQ computer. The system can be customised to 
meet the specific and demanding requirements of the PAU camera. 

\section{Science Capabilities} \label{sec:science} 
\subsection{The BAO scale}\label{sec:BAO}
We have performed extensive simulations of the PAU survey, assuming
8000~deg$^{2}$ covered up to $z=0.9$ using luminous red galaxies with the galaxy density given in Fig.~\ref{fig:nv}
and the redshift precision from Fig.~\ref{fig:res}. This results in measurements
of $H(z)\cdot r_s$ and $d_A(z)/r_s$, $r_s$ being here the sound horizon
at recombination, which are plotted in Fig.~\ref{fig:h-da}. The relative
precision achieved improves monotonically with increasing redshift, flattening
out at about 5\% for $H(z)$ and 2\% for $d_A(z)$.

We have split the redshift interval from $z=0.1$ to $z=0.9$
into 16 equal bins. Results do not change if we change the binning.
Combining these measurements into a cosmology fit, taking into account
correlations, leads to determinations of the parameters $\Omega_{m},w_{0},w_{a}$.
In the following, we will be using the standard parametrization 
\citep{Chevallier2001,linder2003} of
the time evolution of the Dark Energy equation of state, $w(z)=w_{0}+w_{a}\cdot(1-a)$, where $w_0$ denotes the equation of state now, and $w_a$ is (minus) its
current derivative with respect to the scale factor $a$.
The value of the reduced Hubble constant, $h$ drops out of the measured
quantities. The left panel of Fig.~\ref{fig:w0vswa} shows the 68\%
confidence level (CL) contour in the $\Omega_{m}$-$w$ plane that
can be achieved using only PAU LRG data. The corresponding one-sigma
errors are $\sigma(\Omega_m,w) = (0.016,0.115)$.
A flat universe and constant
equation of state has been assumed. In the right panel of Fig.~\ref{fig:w0vswa},
68\% CL contours are shown in the $w_{0}$-$w_{a}$ plane, assuming
a flat universe. The outermost contour approximates the expected world
combined precision from SNe and WMAP when PAU will start taking
data, while the inner contour adds the PAU LRG data to the previous
data set. The reduction in area corresponds to an improvement
by more than a factor three in the DETF figure of merit. The one-sigma
errors are $\sigma(w_0,w_a) = (0.14,0.67)$.

We have also simulated a straw-man spectroscopic survey with equal
area and depth but with $\sigma_{z}=0.0005(1+z)$. The greatly improved
redshift precision results in only a modest $20\%$ decrease in the
cosmological parameter uncertainties. 
Actually, the larger galaxy density that a photometric survey affords
overcompensates for this deficit, as can be seen in Fig.~\ref{fig:compa},
where our simulated reach for several proposed BAO surveys is compared.
Details of the survey characteristics are taken from public
sources and are summarized in Table~\ref{tab:compa}. For Wiggle-Z, we adopt the
total number of galaxies and redshift distribution as given in
\citet{Glazebrook2007}. Surprisingly, our Wiggle-Z constraints on dark energy
parameters are approximately a factor two worse than what one would
predict from the errors on the $d_A$ and $H$ measurements quoted by the
authors. We use the information in section 3 of the SDSS3 project
description (http://www.sdss3.org/collaboration/description.pdf) to
simulate BOSS. The information for HETDEX is taken from \citet{Hill2004}.  
The WFMOS surveys details are taken from the Feasibility Study Report, which can be found in
http://www.gemini.edu/files/docman/science/aspen/.
The Pan-STARRS 1 Survey (PS1) and Dark Energy Survey (DES) information are
taken from their web pages (http://pan-starrs.ifa.hawaii.edu/ and
http://www.darkenergysurvey.org/).
\subsection{Other Probes of Dark Energy} \label{sec:other-probes}
\subsubsection{The galaxy clustering in redshift space}\label{sec:redshift-space}
The redshift accuracy will be sufficiently good to
identify individual structures (walls and voids) along the line of
sight (see Fig.~2). Accurate measurements of the redshift-space power spectrum
will be possible in the linear and mildly non-linear regime, and a
detailed comparison with theoretical predictions will be done in conjunction
with the measurement of Baryon Acoustic Oscillations. Detailed measurements
of redshift space distortions offer an independent test of the growth
history of the peculiar velocity field (i.e.~Newtonian gravitational potential).
This encloses cosmological information on 
dark energy (and/or modified gravity), complementary to that in 
BAO, which measures the background history. A descomposition
of the 2-point correlation function in radial and transverse directions 
also allows for a measurement of bias $b$ and the amplitude
of matter clustering $\sigma_8$, which can be used to study the growth history of
density fluctuations to $z=1$.
Measurement of the amplitude of the galaxy power spectrum, $P(k)$, as
a function of redshift can also be used to determine the growth rate of
structure through the cross-correlation of the galaxy data with 
future CMB lensing data or by using higher-order correlations to determine the bias 
parameter as a function of scale and redshift. Higher-order correlations,
such as the 3-point function or bispectrum, can also be used to measure the BAO feature.
\subsubsection{Weak Lensing}\label{sec:WL}
Weak lensing is sensitive to both the distance and
the growth factor as a function of redshift. The lensing effect can
be measured using either the shear or the magnification.
The PAU camera will not be optimized to measure galaxy
ellipticities, so weak lensing shear may not be as good as those from other
surveys. However, the accurate photometric redshifts obtained in PAU may
be combined with ellipticity measurements obtained in other surveys for the same galaxies. 
This additional information will help separating shear lensing from intrinsic
galaxy aligment.

Gravitational lensing  modulates the observed
spatial distribution of galaxies. Dim galaxies that otherwise would not
have been detected are brought into the sample by the lensing magnification.
This increases the observed number density of galaxies.
On the other hand, magnification also increases the apparent area, which
leads to a drop in the observed number density of galaxies.
The net lensing effect, known as magnification bias,
is controlled by the slope of the number counts. 
The PAU survey will be able to measure this effect
by cross-correlating galaxy samples defined 
by separated redshift slices. 
\subsubsection{Galaxy Clusters}
\label{sec:clusters}
Galaxy clusters are the largest collapsed structures
in the Universe, containing up to hundreds or thousands of individual galaxies.
The redshift distribution and the evolution of
clustering of massive clusters of galaxies can provide a direct
measurement of the cosmic volume 
as a function of redshift as well as the growth rate of density perturbations.
This is complementary to the measurement of the BAO scale, 
which is purely geometrical in nature. Comparison of theory to observations requires 
a calibration of the cluster masses (or at least the mass threshold
 of each cluster sample). Clusters of galaxies can be 
identified optically by searching for concentrations of galaxies with the same color.
The PAU survey by itself will provide a new
window for accurate optical cluster detection and selection, based on the combination
of photometric colors and good photo-z precision over all galaxies
around each cluster, which will help improving on 
cluster completeness and contamination. PAU will also provide the opportunity
to self-calibrate the mass threshold of a given cluster sample in different ways,
such as stacking weak lensing magnification measurements over the cluster positions
or using the (biased) amplitude of clustering in the same cluster sample.
The photo-z accuracy for clusters will be improved
in comparison to the galaxy photo-z by the square
root of the number of galaxies in the cluster. This will result in
a typical photo-z accuracy which is a few times smaller than that for galaxies.
At the same time, one could use the velocity dispersion of the
galaxies in each cluster to provide an estimate of the cluster mass. This will not
be accurate for individual clusters, but should be accurate enough
to have an estimate of the mass threshold of a given cluster sample.

A cluster survey carried out over the PAU area also constrains cosmology through the
spatial clustering of the galaxy clusters. As mentioned above, this can be done
with even higher photo-z accuracy than in the PAU galaxy survey. 
 The clustering of galaxy clusters reflect the underlying clustering in the dark matter;
these correlations contain a wealth of cosmological information, much like the information
contained in the LRG correlation function, including the BAO position. 
Even if the number density for clusters is lower
than that of LRGs, this is partially compensated by the higher (biased)
clustering amplitude. We plan to use the PAU cluster redshift
distribution and the cluster power spectrum and clustering
as cosmological probes to study the density and nature of the dark energy.

PAU can also be used in combination with other
surveys to provide accurate photo-z in a sample of
clusters detected by the  Sunyaev-Zel'dovich (SZ)
or X-ray signatures of hot gas in clusters, as well as in weak 
lensing cluster selection.
\subsection{Other science}\label{sec:other}
The large number of narrow filters in the PAU survey will yield many
colors for all the detected galaxies, allowing the measurement of
numerous interesting parameters for the study of galaxy evolution:
stellar mass, stellar age distribution, metallicity, dust absorption,
and interstellar gas emission. This will make possible a detailed
study of the rates of star formation, galaxy mergers and chemical
evolution that can account for the evolution of the stellar contents
of galaxies of different types, as a function of their environment.

The PAU survey will substantially increase the sensitivity of
astronomical observations to the presence of intergalactic dust, and
possibly detect and characterize it for the first time. Intergalactic
dust extinction can be searched for by correlating the foreground
density of galaxies with the background surface brightness of sources,
as well as the extinction measured from our multiple colors. Our
accurate photometric redshifts will allow for a good estimate of the
mass column density (which should presumably correlate with dust
extinction) in the line of sight to every background galaxy. A detailed
extinction curve for intergalactic dust may be measurable with our
multiple narrow-band colors.  The presence of gray dust 
(causing extinction with no reddening) could also be explored.

The narrow filters of PAU will result in an improved separation of
quasars and stars compared to other surveys, and also a more accurate
estimate of quasar redshifts from the photometry.
Our narrow-band photometry may also be useful
to study the mean transmission of the Lyman alpha forest on large
scales, its fluctuations and its evolution with redshift.
Quasar lens candidates can be searched for by
checking for the presence of multiple images and
followed-up at another observatory with higher angular
resolution. The PAU survey will also provide a map of galaxies in the vicinity of the
lines of sight of all our quasars, with rather accurate photometric
redshifts for all the galaxies. One can follow up with spectroscopic
observations of a selected sample of quasars at another observatory,
and use the PAU galaxies to correlate absorption systems seen in the
quasar spectra with the galaxies, to study the distribution of gas
around galaxies. 

The most luminous star-forming galaxies at redshifts $z \gtrsim 2.5$
will be detectable with the PAU survey narrow-band photometry by means
of their Lyman continuum break, the Lyman alpha forest absorption, and
possibly a Lyman alpha emission line.
This will allow the study of this galaxy population and its clustering
properties over an unprecedentedly large volume.

By selecting the PAU filters appropriately, several
parameters of the stars observed in our survey should be measurable, such as
effective temperature, surface gravity, iron abundance, and $\alpha$/Fe.
An accurate determination of the density profile and metallicity
distribution of halo stars in the Milky Way may follow: the PAU survey
could yield the largest number of metallicity measurements of halo
stars, characterizing the stellar populations of the various streams
believed to have originated our stellar halo.
Giant stars may be also detected by PAU at distances up to $\sim$ 1 Mpc.
This could provide our first substantial sample of stars far from any
galaxy in the Local Group, and extend the measurements of the halo
profile of the Milky Way and its metallicity distributions to much
larger radii. For this purpose, it is necessary to have good ways of
distinguishing nearby K and M dwarfs from distant K and M giants with
the PAU narrow-band photometry. 

PAU should also be great for serendipitous discoveries. We
will have spectral information for every one of the $10^9$ pixels 
in the survey, which will allow for search of diffuse, low signal-to-noise,
components or rare new objects.
For example, if there exist any objects in the universe that produce bursts and
emit most of their power in an emission line, the PAU survey will be
ideal for discovering them.
\section{Summary and Conclusions} \label{sec:summary} 
In 1998, the discovery of the accelerated expansion of the universe 
changed completely our understanding of the universe and its 
components. Ten years on, the quest to understand what causes the 
acceleration continues. Along the way, the Baryon Acoustic Oscillation 
(BAO) technique has been identified as a systematically
robust, yet statistically powerful, probe of dark energy properties.
In particular, measuring the BAO feature along the line of sight as a funcion
of redshift provides a direct determination of $H(z)$, which, in turns, depends on the amount and
characteristics of dark energy. However, such a measurement requires a very precise determination
of galaxy redshifts.

We have presented here a novel approach to photometric redshift determination
which allows the measurement 
of the BAO feature along the line of sight in an efficient way, by using a set of about 40 narrow-band
(FWHM$\approx$~100~\AA) filters. The
approach complements (for BAO and for other science) spectroscopic 
surveys, which typically measure much more precisely the spectra of a much reduced 
sample of galaxies.

Because of the intrinsic width of the peak in the galaxy-galaxy correlation
function of about 15~Mpc/h, there is a fundamental limitation as to how much
one can improve the BAO measurement by reducing the photo-z errors.
A redshift precision of order $\sigma(z)\approx 0.003(1+z)$, corresponding to 15~Mpc/h along the line of sight at $z=0.5$, is about optimal for this measurement. Redshift space distortions, biasing and non-linear effects produce distortions that can be comparable to the effect of this photo-z precision.

Simulations show that both the target galaxy density ($n\sim 10^{-3}$(h/Mpc)$^3$) and precision in redshift
($\sigma_z/(1+z) \sim 0.003$) can be achieved with the proposed system. These simulations indicate that
PAU by itself 
can determine the equation of state of the dark energy assumed constant ($w$) to about
5\%, while when the PAU data is combined with expected supernova and CMB data samples, a sizable increase
in the Dark Energy Task Force figure-of-merit (inverse of the area of the error ellipse in the 
$w_0$--$w_a$ plane)
by about a factor three is achieved, making the PAU very competitive when compared to other planned ground-based
BAO surveys, photometric or spectroscopic.

The survey will produce a unique data set with low-resolution spectroscopy in the optical
wavelengths for all objects in the northern sky up to $m_{AB}=23$--$23.5\ \textrm{arcsec}^{-2}$ (five sigma). 
A survey like PAU, producing
such a catalogue,
will have enormous legacy value and will be extremely useful for many areas of astrophysics,
with contibutions that are different
from, and complementary to, those a spectroscopic survey can deliver. 
\section*{Acknowledgements}
This work was carried out in the framework of the PAU Consolider Collaboration,
supported in part by the 
Spanish Ministry of Education and Science (MEC) through the Consolider 
Ingenio-2010 program, under project CSD2007-00060 ``Physics of the 
Accelerating Universe (PAU).'' Additional support comes from the Barcelona
Supercomputer Center, as well as from the European Commission,
the Spanish High Council for Scientific Research (CSIC), 
and the regional governments of Andalusia, Aragon, Catalonia, Madrid, and Valencia.

\clearpage

\begin{deluxetable}{cccccc}
\tablewidth{0pt}
\tabletypesize{\footnotesize}
\tablecaption{N-body simulations used in this paper. Minimum
halo mass and number of halos correspond to $z=0.5$.
\label{tableN}}
\tablehead{
\colhead{name} & \colhead{$L_{box}$} & \colhead{$N_{par}$}  & 
\colhead{halo mass} & \colhead{$N_{halos}$} \\
\colhead{acronym} & \colhead{$Mpc/h$} & \colhead{number} & 
\colhead{$10^{11} M_{sun}/h$}  & \colhead{Total number} 
}
\startdata
MICE3072 & $3072$ & $1024^3$  & $>375$ & $1.1 \times 10^6$ \\
MICE1536 & $1536$ & $1024^3$  & $>47$  & $2.1 \times 10^6$
\enddata
\end{deluxetable}

\begin{deluxetable}{ccccccccc}
\tablewidth{0pt}
\tabletypesize{\scriptsize}
\tablecaption{Details of the surveys considered in fig.~\ref{fig:compa}.
\label{tab:compa}}
\tablehead{
\colhead{survey}      & \colhead{z range}    & \colhead{Number galaxies} & 
\colhead{Tracer}      & \colhead{Area}       & \colhead{Volume}          & 
\colhead{Radial}      & \colhead{Time scale} & \colhead{reference}      \\
\colhead{}            & \colhead{}           & \colhead{}                &
\colhead{}            & \colhead{deg$^{2}$}  & \colhead{(Gpc/h)$^{3}$}    & 
\colhead{information} & \colhead{}           &\colhead{}
}
\startdata
WiggleZ   & $0.3<z<1.2$ & $2.8\times 10^{5}$ & ELG &  1000 & 2.04 & yes &  2007-2009 & \citet{Glazebrook2007}\\
BOSS-LRG  & $0.2<z<0.8$ & $1.5\times 10^{6}$ & LRG & 10000 & 8.06 & yes &  2009-2014 & see text\\
HETDEX    & $1.8<z<3.7$ & $1.0\times 10^{6}$ & LAE &   200 & 1.91 & yes &    ?       & \citet{Hill2004}\\
WFMOS-ELG & $0.5<z<1.3$ & $2.0\times 10^{6}$ & ELG &  2000 & 4.47 & yes &    ?       & see text\\
WFMOS-LBG & $2.3<z<3.3$ & $6.0\times 10^{5}$ & LBG &   300 & 1.53 & yes &    ?       & see text\\
PS1       & $0.3<z<1.5$ & $5.0\times 10^{8}$ & ALL & 20000 & 65.3 & no  &    ?       & \\
DES       & $0.3<z<1.5$ & $1.5\times 10^{8}$ & ALL &  5000 & 16.3 & no  &  2011-2015 & \\
PAU-LRG   & $0.1<z<0.9$ & $1.3\times 10^{7}$ & LRG &  8000 & 8.6 & yes &  2011-2015 & this paper
\enddata
\end{deluxetable}

\clearpage

\begin{figure}
\epsscale{1}
\plotone{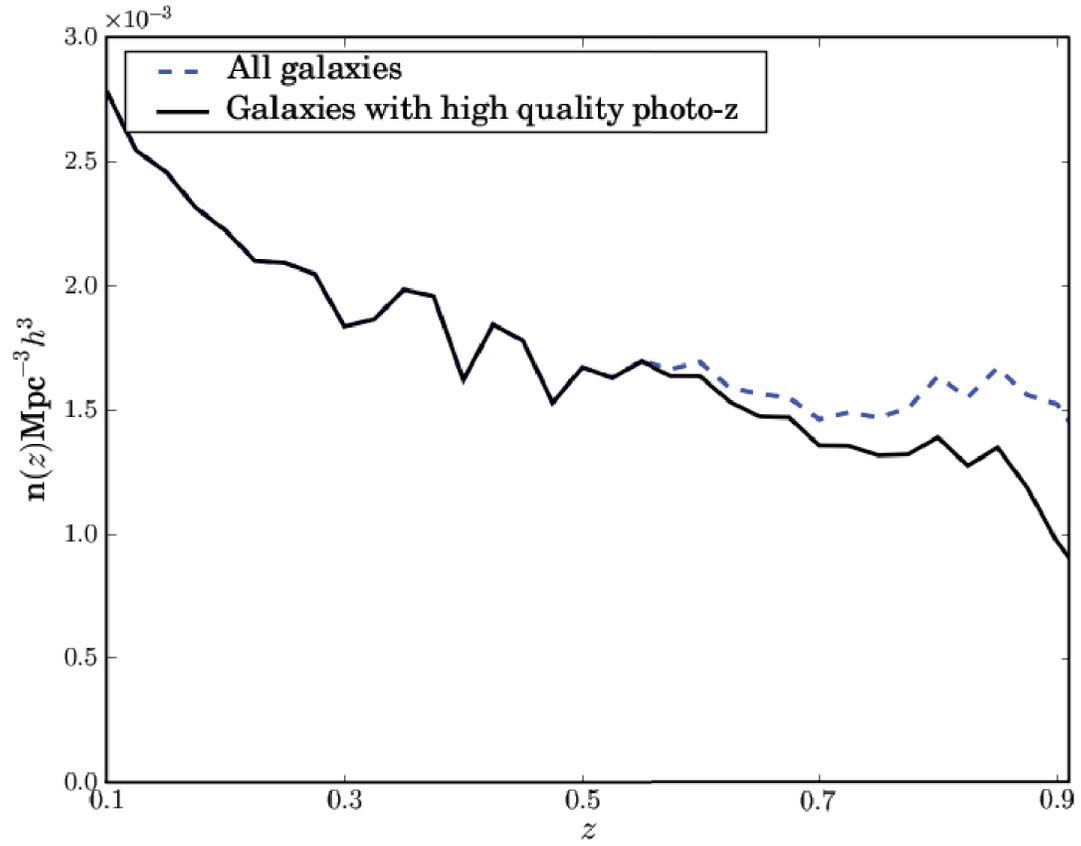}
\figcaption{Spatial density as a function of redshift, for all $L>L_{\star},$$I<23$
red galaxies, and for the subset with high quality photo-z. 
\label{fig:nv}}
\end{figure}

\begin{figure}
\epsscale{1}
\plotone{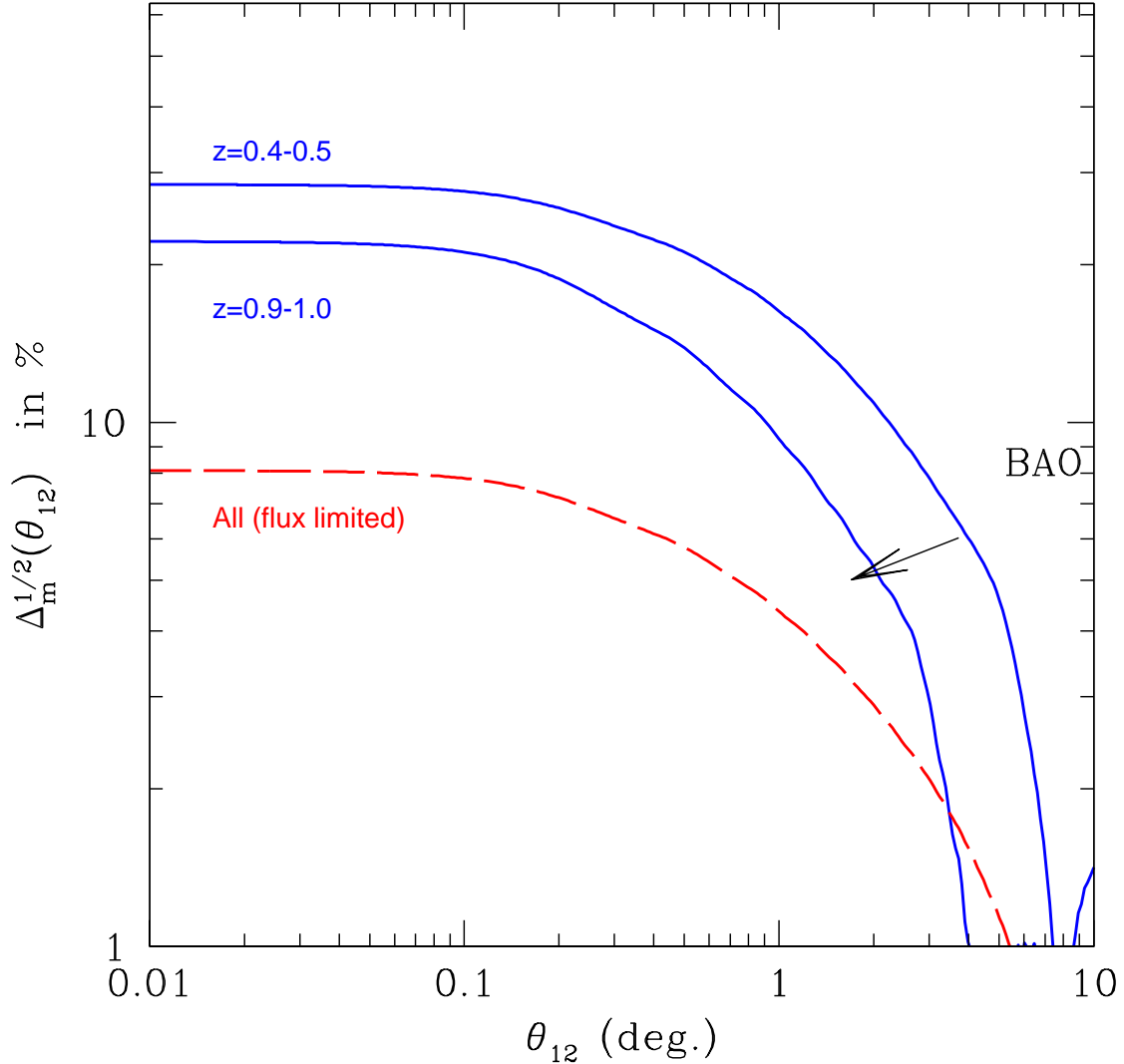}
\figcaption[fig:dm]{Required systematic calibration error (rms percentage) for two
broad redshift slices (thick blue continuous lines): $z=0.4-0.5$
(top) and $z=0.9-1.0$ (bottom) and for a flux
limited sample (red dashed line) including all galaxies to the depth
of PAU (mean $z=0.7$). In these units, at BAO scales (which is a
function of $z$ and is marked by the arrow) the correlation in calibration
error has to be smaller than about $~6\%$ for $z\simeq 0.45$ and
$~5\%$ for $z\simeq0.95$. For other science, the stronger requirements
are driven by the flux limited sample, i.e.~$<2\%$ and $<8\%$ in
correlated errors on scales smaller than $4$ and $0.1$ degrees respectively,
as given by the dashed line.
\label{fig:d_m}}
\end{figure}
   
\begin{figure}
\epsscale{1}
\plottwo{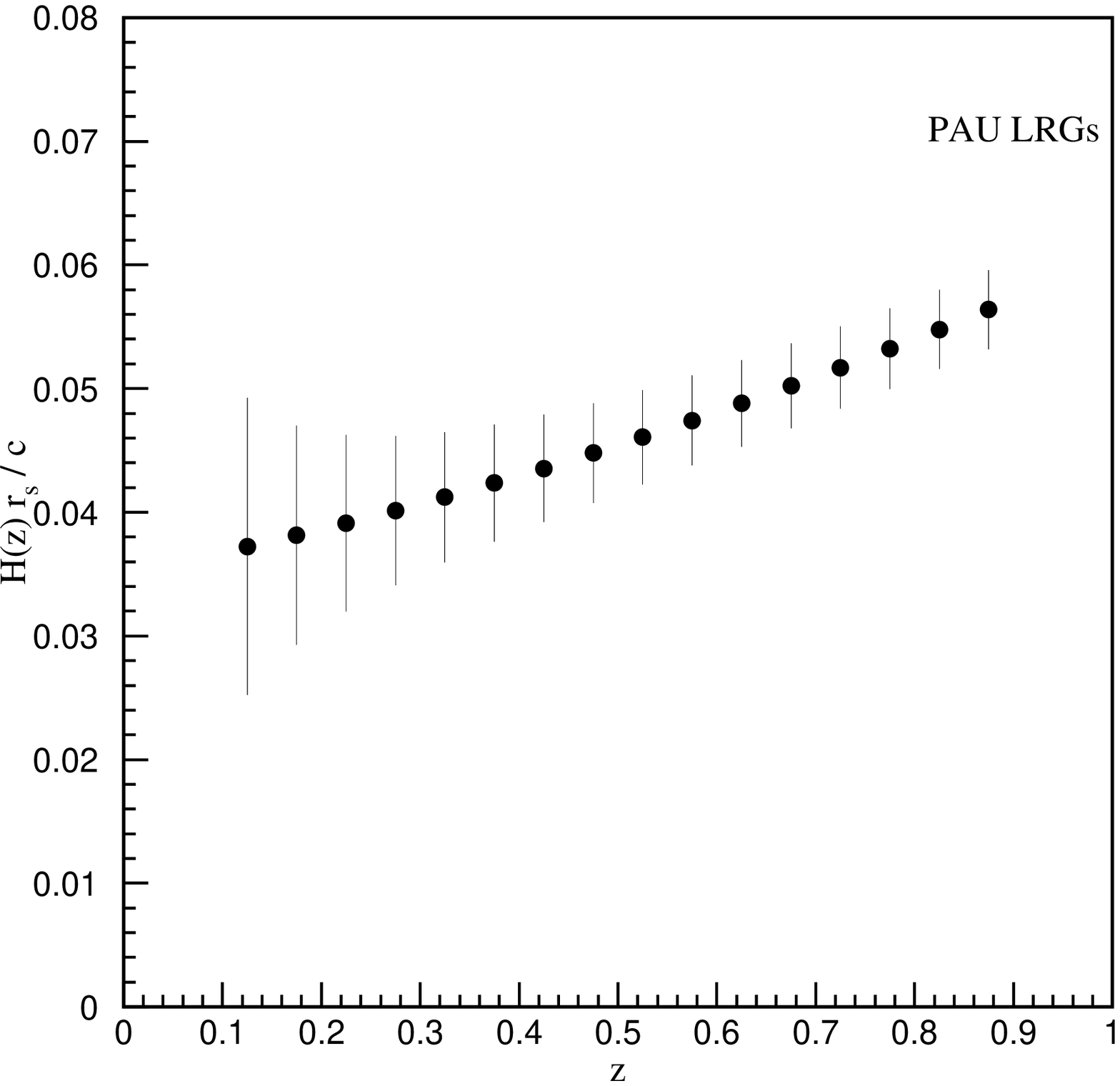}{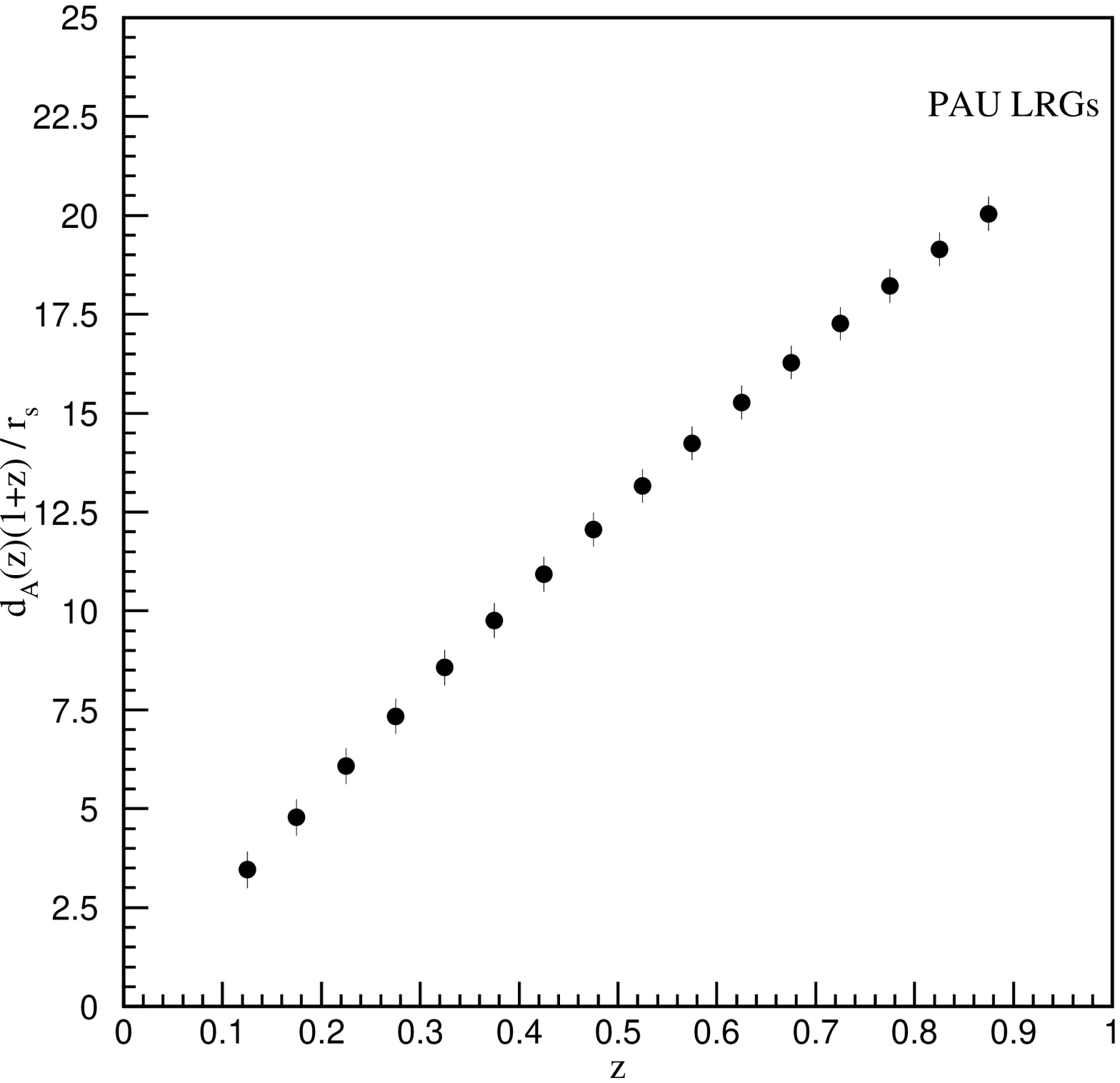} 
\figcaption[h-da]{ \textit{Left:} The expected measurement of radial BAO scale
from the PAU survey (LRGs only).
 \textit{Right:} Same for the measurement of the transverse (angular) BAO scale.
\label{fig:h-da}}
\end{figure}

\begin{figure}
\epsscale{1}
\plottwo{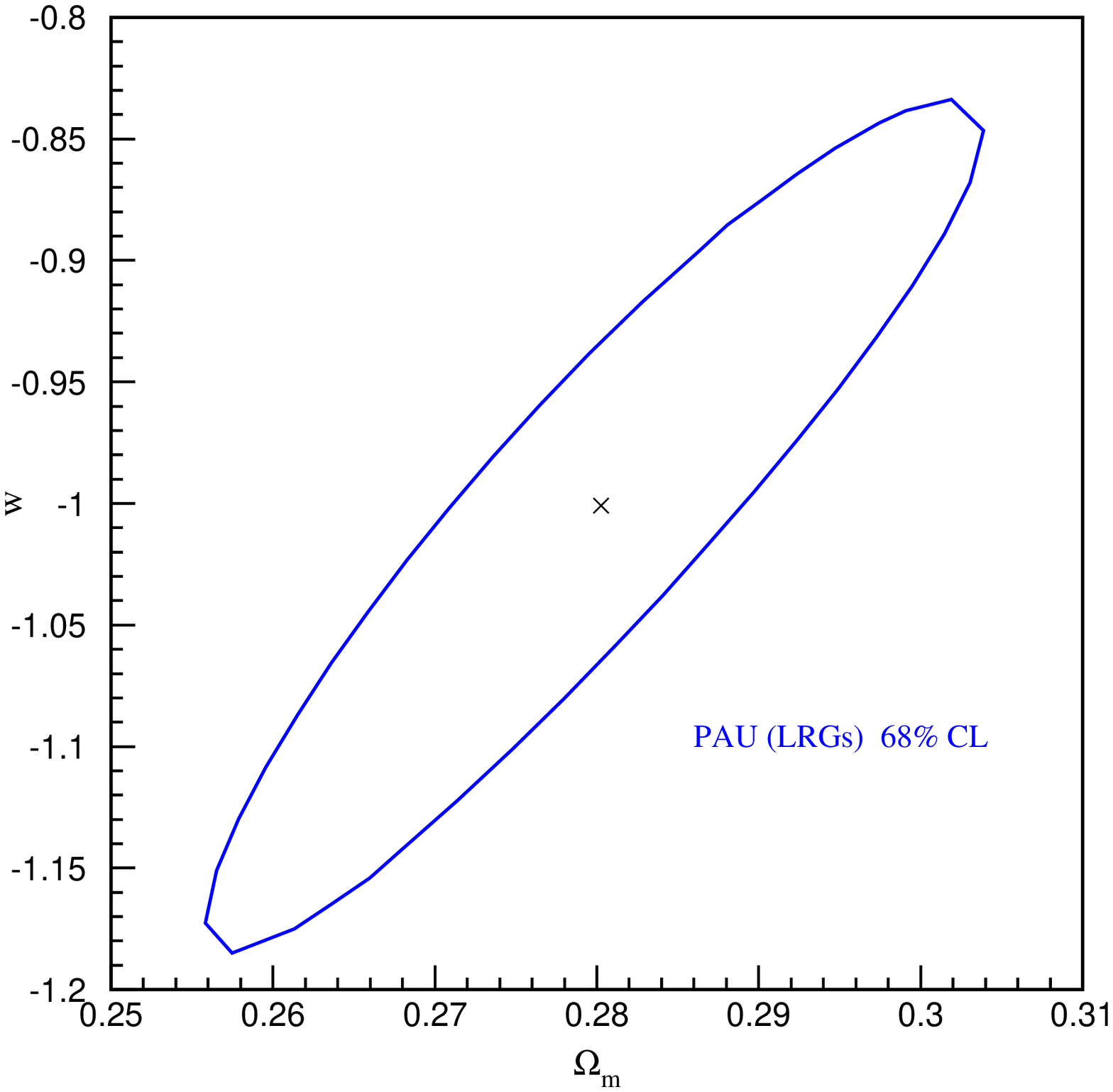}{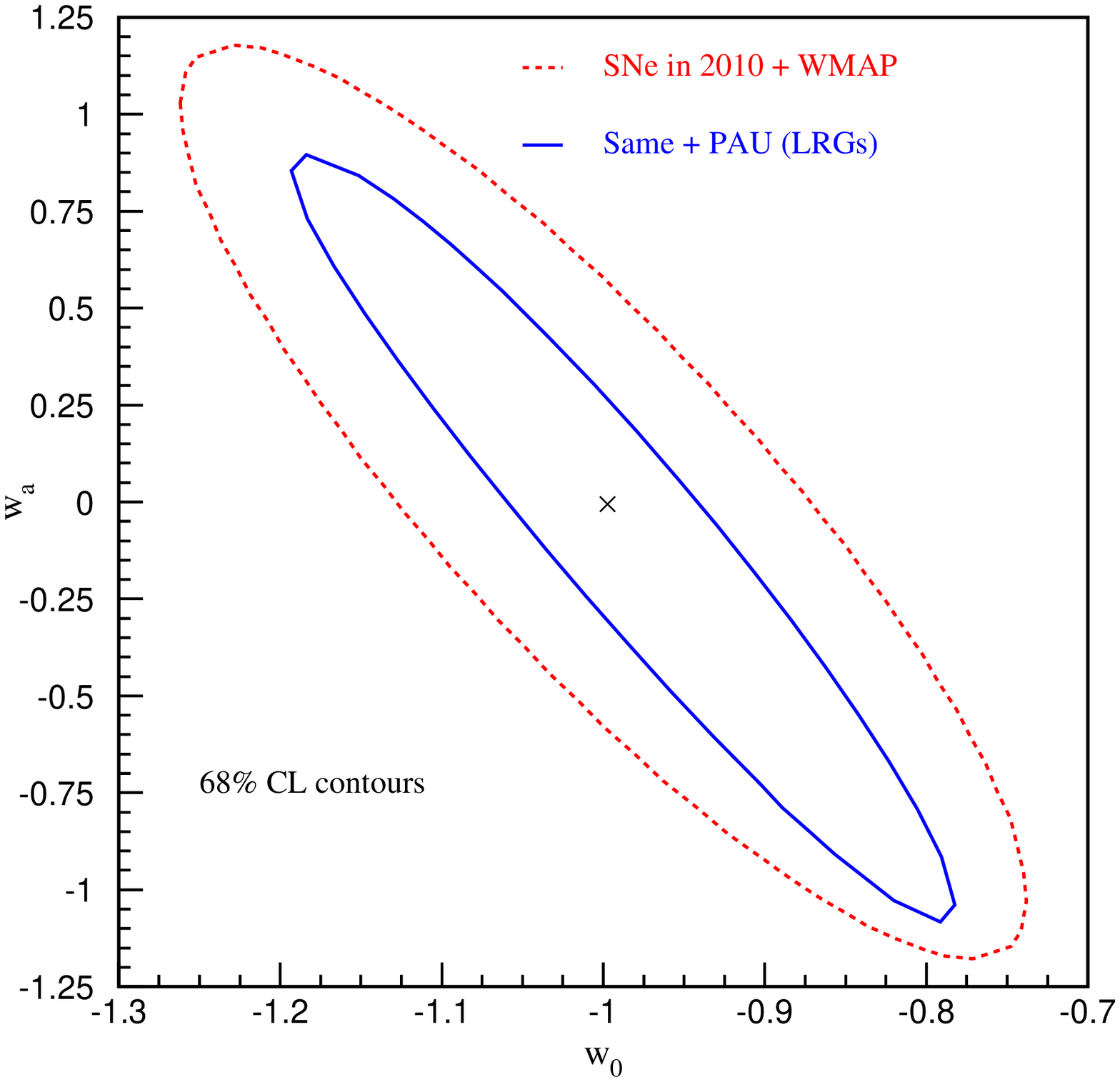} 
\figcaption[w0-wa]{ \textit{Left:} 68\% confidence-level contours in the $\Omega_{m}$-$w$
plane, using only PAU data, assuming a flat universe and a constant
equation of state $w$.
 \textit{Right:} 68\% confidence-level contours in the $w_{0}$-$w_{a}$
plane for the world combined data from SNe and WMAP in about 2010,
and after adding PAU data to that data set. The area of the contour
decreases by about a factor three. 
A flat universe has been assumed. 
\label{fig:w0vswa}}
\end{figure}

\begin{figure}
\epsscale{1}
\plotone{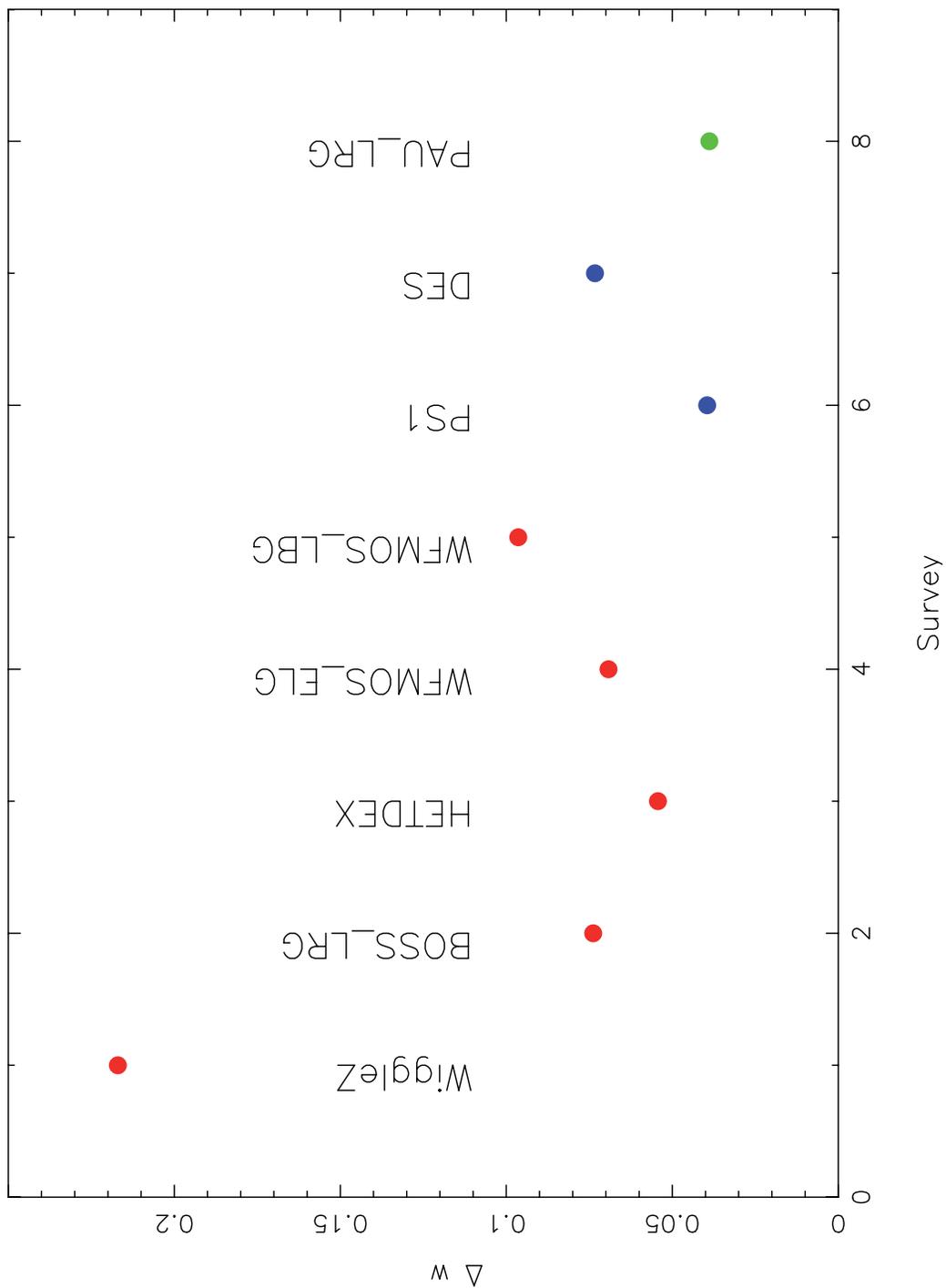} 
\figcaption[Comparison]{Precision on $w$ (assumed constant) for different proposed ground-based BAO surveys. 
All other
cosmological parameters are kept fixed, therefore the overall scale
is unrealistic, but the relative reach of the different proposals
should be realistic. Details about the inputs for the calculations can be found in 
Table~\ref{tab:compa}.
\label{fig:compa}}
\end{figure}

\end{document}